\documentclass[a4paper,11pt]{article}
\pdfoutput=1 

\usepackage{jcappub} 
\usepackage[T1]{fontenc} 

\usepackage{orcidlink}

\usepackage[utf8]{inputenc}
\pdfoutput=1
\usepackage{graphicx}
\usepackage{amsmath}
\usepackage{amsfonts}
\usepackage{amssymb}
\usepackage{xcolor,soul}
\usepackage{epstopdf}
\usepackage{xcolor}
\usepackage{csquotes}

\usepackage{float}
\usepackage{subcaption}

\usepackage{relsize}
\usepackage{graphics}
\usepackage{hyperref}
\usepackage{mathrsfs}
\usepackage{amssymb}
\usepackage{booktabs}
\usepackage[normalem]{ulem}
\usepackage{amsthm}
\usepackage{cancel}
\usepackage{fontawesome}   
\usepackage{tabularx,ragged2e}

\usepackage{tikz}
\usetikzlibrary{positioning,decorations.pathmorphing}

\hypersetup{
     colorlinks   = true,
     citecolor    = blue,
     linkcolor    = blue,
     urlcolor     = blue,
}
\usepackage{physics}

\title{
Revisiting PBH accretion, evaporation and their cosmological consequences}

\author{Jitumani Kalita\,\orcidlink{0009-0001-5681-6194}}
\author{and Debaprasad Maity\,\orcidlink{0000-0002-5458-7121}}

\affiliation{Department of Physics, Indian Institute of Technology, Guwahati, 
Assam, India}

\emailAdd{k.jitumani@iitg.ac.in}
\emailAdd{debu@iitg.ac.in}

\abstract{Primordial black holes (PBHs) provide a unique probe of the early Universe. Their cosmological evolution is governed by the competition between mass accretion and Hawking evaporation. In this paper we look into the details impact of accretion. Most of the earlier analysis relied on non-relativistic accretion models. In this work, we reinvestigate this in a fully relativistic framework for Kerr PBHs in the radiation-dominated era. We derive relativistic accretion rate and compute spin-dependent efficiency $\lambda_{\text{Kerr}}(a_*)$. Using this result, we construct coupled evolution equations for the PBH mass and spin that include both relativistic accretion and spin-dependent evaporation. Our analysis shows that relativistic accretion significantly increases PBH masses and consequently suppresses their spins, causing all PBHs to become effectively Schwarzschild well before evaporation. These effects strengthen the Big Bang Nucleosynthesis (BBN) bound on the initial PBH mass by a factor of $\sim 4$--$5$, reduce the mass required for survival to the present epoch to $\sim 2.7\times 10^{14}\,\mathrm{g}$, and shift the viable particle like DM parameter space. Notably the early accretion induced spin-down effect further washes out the well known high-frequency, spin-induced feature in the high frequency stochastic gravitational-wave background, modifying predictions for future detectors.}

\keywords{Accretion, Primordial Black Hole, Dark Matter, Gravitational Wave}

\newcommand{\bea}{\begin{aligned}}
\newcommand{\eea}{\end{aligned}}

\def\beq{\begin{equation}}
\def\eeq{\end{equation}}

\def\beqa{\begin{eqnarray}}
\def\eeqa{\end{eqnarray}}

\def\be{\begin{equation}}
\def\ee{\end{equation}}

\def\bse{\begin{subequations}}
\def\ese{\end{subequations}}

\def\bea{\begin{eqnarray}}
\def\eea{\end{eqnarray}}

\def\Min{M_{\mathrm{in}}}

\usepackage{pgfplots}
\pgfplotsset{compat=1.17}

\begin{document}
\maketitle
\flushbottom

\section{Introduction}\label{intro}

Primordial Black Holes (PBHs) remain one of the most compelling and consequential theoretical possibilities in modern cosmology \cite{Carr:1974nx, Hawking:1971ei, Zeldovich:1967lct, Green:2014faa, Dai:2024guo}. Formed from the collapse of large density fluctuations in the very early universe~\cite{Young:2019yug, Jedamzik:1996mr, Kawaguchi:2007fz, Kim:1999xg, Yoo:2024lhp, Yoo:2022mzl, Escriva:2021aeh}, PBHs are a unique probe of high-energy physics, the inflationary epoch, and the properties of the primordial power spectrum \cite{Sasaki:2018dmp, Carr:2020gox, Carr:2009jm, Sasaki:2016jop, Clesse:2016vqa}. Their predicted mass function is broad, spanning dozens of orders of magnitude, and their existence is neither confirmed nor entirely ruled out~\cite{Frampton:2015png, Kirillov:2021qjz, Khlopov:2008qy, DeLuca:2020ioi}. This vast parameter space allows PBHs to play a multitude of roles in cosmic history. In the high-mass regime ($M \gtrsim 10^{15} \text{ g}$), they are a viable, non-particulate candidate for the entirety of the universe's dark matter (DM) \cite{Bird:2016dcv, Carr:2016drx}. In the low-mass regime ($M < 10^{15} \text{ g}$), PBHs are unstable and would have evaporated by the present day due to Hawking radiation \cite{Hawking:1975vcx}. Although no longer present, these light PBHs could have profoundly impacted cosmic evolution. Their evaporation can inject high-energy particles into the primordial plasma, altering the abundances of light elements synthesized during Big Bang Nucleosynthesis (BBN) \cite{Carr:2009jm, Keith:2020jww}. Furthermore, if they evaporate to produce stable, beyond Standard Model particles, PBHs themselves could be the source of the DM relic abundance observed today \cite{Gondolo:2020uqv, Barman:2024iht, Khan:2025kag}.

The viability of all these scenarios depends critically on one thing: a precise understanding of the PBH mass evolution, $M(t)$. This evolution is a competition between two primary processes: mass loss from Hawking evaporation and mass gain from the accretion of the surrounding cosmic fluid. For decades, the study of these two processes has been largely bifurcated. On one hand, Hawking evaporation has been studied in great detail, with sophisticated calculations of greybody factors for rotating black holes (BHs) \cite{Page:1976df, Cvetic:1997xv, Jorge:2014kra, Konoplya:2024vuj, Noda:2022zgk} and, models of PBH accretion have, until recently, relied on the oversimplified Bondi (or Bondi-Hoyle-Lyttleton) accretion formula \cite{Bondi:1952ni}. This non-relativistic approximation, derived for a static BH in a non-expanding medium, is ill-suited for its dynamic in an environment such as radiation-dominated universe. This simplification is problematic for two main reasons. First, the fluid in the early universe is a relativistic plasma ($p = \rho/3$), and the gravitational field near the BH is inherently relativistic. A fully general-relativistic treatment of the fluid dynamics is therefore inevitable. Second, the initial spin of PBHs depends heavily on their formation mechanism. While standard formation from the collapse of density perturbations in a radiation-dominated universe yields PBHs with very small initial spins ($a_* \lesssim 0.01$)~\cite{DeLuca:2019buf}, alternative scenarios---such as collapse during an early matter-dominated phase---allow PBHs to form with non-zero, and potentially maximal, spin \cite{Carr:1976zz, Mirbabayi:2019uph, Harada:2017fjm}. Regardless of the initial conditions, the spacetime of a spinning (Kerr) BH fundamentally alters the fluid's in-fall dynamics due to general relativistic corrections to the effective potential and the capture cross-section.


A complete and self-consistent model that simultaneously incorporates \textit{relativistic accretion} onto a \textit{Kerr} BH, embedded in a cosmological background, and its competition with spin-dependent \textit{Hawking evaporation} has been a significant gap in the literature. It is this gap that we aim to fill. In this paper, we construct a complete and self-consistent model for the evolution of both the mass $M(t)$ and spin $a_*(t)$ of a PBH, from its formation in the radiation era to its final evaporation. Our method proceeds as follows. We first derive the full general-relativistic hydrodynamic equations for a perfect fluid in a general, stationary, and axisymmetric spacetime. We then specialize this to the Kerr metric, assuming a steady-state, radially in-falling fluid with zero angular momentum as an  expected condition for the ambient, homogeneous cosmic fluid. By solving the flow equations, we numerically determine the critical (sonic) point and derive the relativistic accretion rate. This rate is parameterized by a new dimensionless efficiency $\lambda_{\text{Kerr}}(a_*)$, which serves as the general-relativistic correction to the non-relativistic Bondi formula. We then combine this new accretion rate with the well-established formalism for Hawking evaporation, using the precise evaporation rates from the public codes \texttt{BlackHawk} \cite{Arbey:2019mbc, Arbey:2021mbl} and \texttt{FRISBHEE} \cite{Cheek:2021odj, Cheek:2021cfe}, which include the full greybody factors. This results in a coupled system of differential equations for $dM/dt$ and $da_*/dt$, which we solve numerically.

It is important to clearly distinguish our focus on long-term evolutionary dynamics from numerical studies of PBH formation. State-of-the-art general relativistic simulations~\cite{deJong:2021bbo, Musco:2023dak, Musco:2004ak, Escriva:2019phb} rigorously track the non-linear collapse of overdensities and the immediate post-formation fluid dynamics, including formation during the radiation-dominated era. In contrast, we performed semi-analytical analysis of an already-formed PBH living in radiation bath with naturally large over density threshold. Furthermore, while the simulation results provide the crucial initial conditions for the PBH population, our analysis essentially captures their subsequent cosmological trajectory. By incorporating BH spin, a parameter largely absent in formation-focused simulations, we capture the long-term competition between relativistic accretion and spin-dependent Hawking evaporation up to cosmological timescales.

Our results reveal several novel physical effects. We find that relativistic accretion is significantly more efficient than its non-relativistic counterpart, leading to a substantial increase in the PBH mass before evaporation takes over. This accretion is strongly suppressed by the BH's spin, with $\lambda_{\text{Kerr}}(a_*)$ decreasing for higher $a_*$. Most critically, we find that the accretion of the zero-angular-momentum fluid acts as a source of ``spin dilution''. This effect rapidly spins down the BH, causing even a maximally-spinning ($a_* \approx 0.99$) PBH to evolve into an effectively non-rotating ($a_* \approx 0$) Schwarzschild BH long before it begins its final, violent evaporation.

These physical effects have profound consequences for all cosmological constraints on PBHs. First, for light, evaporating PBHs, the accretion-induced mass gain dramatically extends their lifetime ($\tau_{\text{evap}} \propto M^3$). This means a PBH of a given initial mass evaporates much later than previously thought. As a result, the BBN constraint (which demands evaporation \textit{before} $T \sim 4 \text{ MeV}$~\cite{Hannestad:2004px}) becomes significantly stronger, ruling out initial masses a factor of $\sim 4-5$ smaller than the standard limit. For the same reason, PBHs that produce DM now have a longer lifetime to do so, strengthening the constraints on their initial abundance, $\beta$. Second, for heavier, DM candidate PBHs, this mass growth implies that the critical initial mass required for a PBH to survive to the present day is lowered from the canonical $M_{\text{in}} \sim 10^{15} \text{ g}$~\cite{Carr:2009jm} to a new limit of $M_{\text{in}} \sim 2.7 \times 10^{14} \text{ g}$. This fundamentally shifts the entire $f_{\text{PBH}}$ parameter space. Observational constraints from extragalactic evaporation, microlensing, and gravitational wave (GW) mergers, which are all sensitive to the \textit{present-day} mass $M_0$, are now mapped onto a new, smaller range of \textit{initial} masses. Finally, the spin-down effect has a smoking-gun observational consequence for the stochastic gravitational wave background (SGWB)~\cite{Romano:2016dpx, Caprini:2018mtu, Yuan:2021xdi, Dong:2015yjs}. The predicted high-frequency ``bump'' in the GW spectrum, a unique signature of high-spin PBHs, is completely erased. Our model predicts that all accreting PBHs, regardless of their initial spin, evaporate as Schwarzschild BHs, producing a smooth, single-peaked GW spectrum. However, note that in the analysis the accreting fluid is assumed to have negligible initial angular momentum, which seems reasonable in the radiation dominated universe.

This paper is organized as follows. In Sec.~\ref{sec:background}, we derive the general relativistic
hydrodynamic equations governing accretion in a stationary, axisymmetric spacetime. In Sec.~\ref{sec:mass_acc}, we apply this framework to the Kerr geometry embedded in an expanding
FRW universe and numerically determine the spin--dependent accretion efficiency $\lambda_{\text{Kerr}}(a_*)$. In Sec.~\ref{sec:mass_evp}, we review the formalism for Hawking evaporation and black–hole spin-down. In Sec.~\ref{sec:mass_acc_evp}, we present the full set of coupled evolution equations for $M(t)$ and $a_*(t)$, along with their numerical solutions. In Secs.~\ref{sec:BBN}, \ref{sec:DM}, and \ref{sec:f_PBH}, we derive the updated cosmological constraints arising from BBN, DM production, and the resulting bounds on $f_{\text{PBH}}$, respectively. In Sec.~\ref{sec:GW}, we provide the corresponding predictions for the modified SGWB spectrum. Finally, we summarize our results in Sec.~\ref{sec:conclusion}.


\section{Relativistic Hydrodynamics in Axisymmetric Spacetime}\label{sec:background}

Consider a general axisymmetric spacetime, whose line element can be expressed as~\cite{Konoplya:2016jvv, Harmark:2005vn}
\begin{align}
ds^2 = g_{\mu \nu} dx^{\mu} dx^{\nu} = g_{tt} dt^2 + 2 g_{t\phi} dt\, d\phi + g_{rr} dr^2 + g_{\theta \theta} d\theta^2 + g_{\phi \phi} d\phi^2,
\end{align}
where the indices $\mu, \nu$ run over the coordinates $(t, r, \theta, \phi)$. We assume the BH is located at $r=0$ with the event horizon defined by the condition $g^{rr} = 1/g_{rr} = 0$. In general, all metric components are functions of the coordinates $(r, \theta)$ only, reflecting the axial symmetry and stationarity of the spacetime. The dynamics of the accreting fluid are governed by the conservation of the stress-energy tensor $T^{\mu\nu}$ and the particle number current $j^{\mu}$
\begin{equation}
\nabla_{\nu} T^{\mu\nu}=0 \quad \text{and} \quad \nabla_{\mu} j^{\mu}=0.
\end{equation}
For a perfect fluid, these are given by
\begin{equation}
\begin{aligned}
    T^{\mu\nu} &= (\rho+p)u^{\mu}u^{\nu}+pg^{\mu\nu}, \\
    j^{\mu} &= \tilde{\rho} u^{\mu},
\end{aligned}
\end{equation}
where $\rho$ is the total energy density, $p$ is the pressure, $\tilde{\rho}$ is the rest mass density, and $u^{\mu}$ is the fluid four-velocity, satisfying $u_{\mu}u^{\mu}=-1$. To derive the equation of motion, we project the energy-momentum conservation equation onto the hypersurface orthogonal to the fluid's four-velocity using the projection operator $h^i_{\mu} = \delta^i_{\mu} + u^i u_{\mu}$. This procedure, $h^i_{\mu}\nabla_{\nu}T^{\mu\nu}=0$, yields the relativistic Euler equation~\cite{Dihingia:2018tlr, Das:2025vts}
\begin{equation}
\label{eq_euler}
(\rho+p)u^{\nu}\nabla_{\nu}u^{i} + (g^{i\nu}+u^i u^{\nu})\partial_{\nu}p = 0 ,
\end{equation}
where $i$ is a spatial index. This equation describes how the fluid is accelerated by pressure gradients. Furthermore, projecting the conservation law along the flow itself, $u_{\mu}\nabla_{\nu}T^{\mu\nu}=0$, and combining it with the continuity equation $\nabla_{\mu}j^{\mu}=0$, we obtain a scalar equation relating the thermodynamic quantities
\begin{equation}
\label{eq_thermo}
u^{\mu}\left[\left(\frac{\rho+p}{\tilde{\rho}}\right)\partial_{\mu}\tilde{\rho} - \partial_{\mu}\rho\right] = 0.
\end{equation}
We introduce the following variables to parameterize the three-velocity components in general axisymmetric spacetimes: Azimuthal (angular) velocity $
v_{\phi}^2 = \frac{u^{\phi} u_{\phi}}{ - u^t u_t },
$ with its corresponding Lorentz factor given by $\gamma_{\phi}^2 = (1 - v_{\phi}^2)^{-1}$. The polar (latitudinal) velocity component is then defined as $
v_{\theta}^2 = \gamma_{\phi}^2\, \frac{u^{\theta} u_{\theta}}{ - u^t u_t }, $ which has an associated Lorentz factor of $\gamma_{\theta}^2 = (1 - v_{\theta}^2)^{-1}$. Similarly, the square of the radial velocity is
$v_{r}^2 = \frac{u^r u_r}{ - u^t u_t }$. The total Lorentz factor, $\gamma$, is constructed as a product of these components, $\gamma = \gamma_{\phi} \gamma_{\theta} \gamma_{v}$. The final Lorentz factor, $\gamma_v$, is defined by $\gamma_v^2 = (1-v^2)^{-1}$, where the velocity $v$ is a composite of the radial velocity and the other Lorentz factors, such that $v^2 = \gamma_{\phi}^2 \gamma_{\theta}^2 v_r^2$.

For this study, we consider a steady-state ($\partial_t=0$), axisymmetric ($\partial_\phi=0$) flow. We make the crucial simplification that the accretion flow is \textit{purely radial} with no motion in the polar direction ($u^{\theta}=0$). We will be applying our formalism in the early radiation dominated universe. During the radiation dominated phase large positive pressure tends to erase the local inhomogeneity. Therefore, while undergoing accretion, it is reasonable to assume that the fluid elements start their journey with negligible azimuthal velocity $u_\phi$. Hence, we assume the fluid has \textit{zero angular momentum} ($u_{\phi}=0$) through out. In the rotating Kerr spacetime, however, this does not imply the coordinate angular velocity is zero; due to frame-dragging, the fluid must acquire a non-zero $u^{\phi} = - (g_{t\phi}/g_{\phi\phi})u^t$. Under these assumptions, the $i=r$ component of the Euler equation Eq.~(\ref{eq_euler}) becomes
\begin{align}
(\rho+p)(u^{t}\nabla_{t}u^{r}+u^{r}\nabla_{r}u^{r}+u^{\phi}\nabla_{\phi}u^{r})+(g^{rr}+u^r u^{r})(\partial_{r}p)=0 .
\end{align}
After substituting the relevant Christoffel symbols for the general metric and simplifying using the normalization condition $u_\mu u^\mu = -1$, we arrive at the final dynamical equation for the radial velocity $v$ as
\begin{equation}
\label{eq_velocity_final}
\gamma^2 v \frac{dv}{dr} + \frac{d\Phi^{\text{eff}}}{dr} + \frac{1}{\rho+p}\frac{dp}{dr}=0 ,
\end{equation}
where $v$ is the radial three-velocity, $\gamma=(1-v^2)^{-1/2}$ is the corresponding Lorentz factor, and $\Phi^{\text{eff}}$ is the effective potential experienced by the fluid particles, given by
\begin{equation}\label{eq_phi_eff}
\Phi^{\text{eff}} = \frac{1}{2}\ln \left( \frac{g_{t\phi}^2}{g_{\phi\phi}} - g_{tt} \right).
\end{equation}
Finally, integrating the continuity equation $\nabla_{\mu}j^{\mu}=0$ over a spherical surface gives the total mass accretion rate as
\begin{equation}
\label{eq_mdot_final}
\dot{M} = - \oint \sqrt{-g} \, j^r \, d\theta d\phi = 2\pi \int_0^{\pi} \tilde{\rho} \, \gamma v \sqrt{g^{rr}}\sqrt{-g} \, d\theta .
\end{equation}
These two equations, Eq.~(\ref{eq_velocity_final}) and Eq.~(\ref{eq_mdot_final}), form the basis of our analysis.

\subsection{Application to Kerr Spacetime}

\begin{figure}[t]
\centering
\includegraphics[width=.497\textwidth]{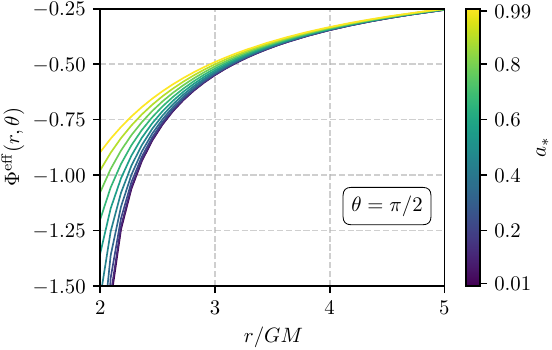}
\hspace{0.01cm}
\includegraphics[width=.48\textwidth]{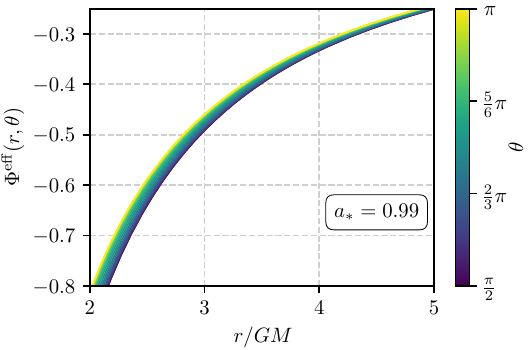}
\caption{The effective potential $\Phi^{\text{eff}}(r, \theta)$ as a function of the dimensionless radial distance $r/GM$ is plotted. \textbf{Left Panel:} The potential in the equatorial plane ($\theta = \pi/2$) for various spin parameters $a_*$, as indicated by the color bar. \textbf{Right Panel:} The potential at a fixed, high spin ($a_* = 0.99$) for different polar angles $\theta$, from the equator ($\theta = \pi/2$) to the pole.}
\label{Two_plots_phi_eff}
\end{figure}

We now specialize our general formalism to the case of accretion onto a rotating Kerr BH. In Boyer-Lindquist coordinates, the Kerr metric is defined by the components~\cite{PhysRevLett.11.237, Boyer:1966qh}
\begin{equation}
\begin{aligned}
    g_{tt} &= -\left(1-\frac{r_g r}{\Sigma}\right), \quad g_{rr} = \frac{\Sigma}{\Delta}, \quad g_{\theta\theta} = \Sigma,  \\
    g_{\phi\phi} &= \frac{A \sin^2\theta}{\Sigma}, \quad g_{t\phi} = -\frac{a_k r_g r\sin^2\theta}{\Sigma} ,
\end{aligned}
\end{equation}
where we have used the standard definitions: $r_g=2GM$, $\Sigma = r^2 + a_k^2\cos^2\theta $, $\Delta = r^2 - r_g r + a_k^2 $ and $A = (r^2+a_k^2)^2 - \Delta a_k^2\sin^2\theta $. Here, $M$ is the BH mass, $J$ is the angular momentum and $a_k=J/M$ is the specific spin parameter. The outer event horizon is located at the larger root of $\Delta=0$, which is $r_+ = GM(1 + \sqrt{1 - a_*^2})$, with $a_* = J/GM^2$ is the dimensionless Kerr parameter. The determinant of the metric is $\sqrt{-g} = \Sigma\sin\theta$. Substituting the relevant metric components into the general accretion rate formula, Eq.~(\ref{eq_mdot_final}), yields the accretion rate for the Kerr spacetime 
\begin{equation}
\label{eq_kerr_mdot}
\dot{M} = 2\pi \int_0^{\pi} \tilde{\rho} \, \gamma v \sqrt{\Delta\Sigma} \sin\theta \, d\theta .
\end{equation}
This integral explicitly shows the dependence of the accretion rate on the BH's mass $M$, spin $a$, and the radial and polar distribution of the surrounding fluid.

Similarly, the effective potential governing the fluid's radial acceleration, from Eq.~(\ref{eq_phi_eff}), becomes
\begin{equation}
\label{eq_kerr_potential}
\Phi^{\text{eff}}(r, \theta) = \frac{1}{2}\ln\left(\frac{a_k^2 r^2 r_g^2 \sin^2\theta}{\Sigma A} + 1 - \frac{r_g r}{\Sigma}\right).
\end{equation}
The effective potential is shown in Fig.~\ref{Two_plots_phi_eff} as a function of the dimensionless radial distance $r/GM$. This shows, as the spin $a_*$ increases, the potential becomes less negative, indicating a stronger repulsive barrier due to the frame-dragging effect.
The gradient of this potential, along with the pressure gradient, drives the fluid's infall. The set of equations is now fully specified for the Kerr geometry and ready for numerical or analytical solution.

\section{PBH Mass Evolution: Accretion}\label{sec:mass_acc}

To apply our formalism to PBHs in the early universe, we must embed the accretion process within an expanding cosmological background. We model the cosmic fluid with a simple barotropic equation of state and solve the hydrodynamical equations by assuming that the fluid follows the Hubble flow at large distances from the BH.

\subsection{Thermodynamics and Equation of State}

We assume the background cosmic fluid is a perfect fluid with the equation of state $p = \omega \rho$, where for the radiation-dominated era, $\omega = 1/3$~\cite{Lozanov:2016hid}. The relationship between the rest mass density $\tilde{\rho}$ and the energy density $\rho$ is given by the thermodynamic relation derived from Eq.~(\ref{eq_thermo}) as
\begin{equation}
    \frac{d\rho}{\rho} = (1+\omega) \frac{d\tilde{\rho}}{\tilde{\rho}} .
\end{equation}
Integrating this gives a direct relation between the two densities
\begin{equation}
\label{eq:rho_relation}
\tilde{\rho} = \rho_{\infty}^{\frac{\omega}{1+\omega}} \rho^{\frac{1}{1+\omega}} ,
\end{equation}
where the integration constant has been set by the boundary condition that far from the PBH, the fluid is at rest in the comoving frame, with energy density equal to the background cosmic value, $\rho = \tilde{\rho} = \rho_{\infty}(t)$.

\subsection{The Cosmological Flow Solution}


The effect of cosmic expansion is taken into account by expressing the radial coordinate as $r = a(t) x$. In this coordinate system, the radial velocity $v(r)$ can be written in terms of the comoving coordinate $x$ and the peculiar velocity $v_p$ as
\begin{equation}
    v(r) = H(t)r + v_p.
\end{equation}
It is important to assess the validity of employing an asymptotically flat Kerr metric within an expanding background. A fully self-consistent treatment would require a dynamical metric describing a rotating BH embedded in an FLRW universe. To justify our patched approximation, it is important to analyze the hierarchy of the relevant scales. At the time of PBH formation, the BH horizon radius $ r_+ $ is related to the Hubble horizon $ H^{-1} $ through $ r_+ \simeq 0.1\,H^{-1} $. However, the rapid expansion of the Universe quickly separates these scales. Soon after formation, the condition $ r_+ \ll H^{-1} $ becomes well satisfied, implying $ \eta \equiv r_+ H \ll 1 $. Consequently, we can define a characteristic length scale, $r_*$, which denotes the BH's sphere of influence where its local gravity dominates over the background cosmic expansion. This radius can be approximated by equating the BH mass $M$ to the total background fluid mass contained within a volume of radius $r_*$, such that $M \simeq \frac{4}{3}\pi r_*^3 \rho$. This relationship yields $r_* \simeq r_+ / \eta^{2/3}$. For small $\eta$, the BH's gravitational influence extends significantly beyond its horizon ($r_* \gg r_s$). Within this intermediate region ($r_+ < r \ll r_*$), the fluid dynamics are primarily governed by the quasi-static Kerr geometry, and the Hubble flow can be safely treated as a small perturbation to the local in-fall velocity. At much larger distances ($r > r_*$), the background fluid transitions to being completely described by the standard cosmological expansion dictated by the scale factor $a(t)$.

To determine the density profile of the accreting gas, we integrate the radial momentum equation, Eq.~(\ref{eq_velocity_final}), from a large distance (where $\rho = \rho_{\infty}$ and $v_p \approx -Hr$) down to a radius $r$. This yields
\begin{equation}
    \int_{0}^{v} \frac{v' dv'}{1-(v')^2} + \int_{\rho_{\infty}}^{\rho} \frac{\omega}{(1+\omega)} \frac{d\rho'}{\rho'} + \int_{\infty}^{r} d\Phi^{\text{eff}} = 0 .
\end{equation}
Solving this integral gives the local fluid density $\rho$ as a function of the local velocity $v$ and the effective potential
\begin{equation}
\label{eq:density_profile}
\rho(r,\theta) = \rho_{\infty} \left[ 1 - v(r)^2 \right]^{\frac{1+\omega}{2\omega}} \left[ \mathcal{F}(r, \theta, a_*) \right]^{-\frac{1+\omega}{2\omega}} ,
\end{equation}
where we have defined $\mathcal{F} \equiv \exp(2\Phi^{\text{eff}})$ for compactness, representing the full geometric influence of the Kerr spacetime.

Finally, we substitute the density profile from Eq.~(\ref{eq:density_profile}) and the thermodynamic relation from Eq.~(\ref{eq:rho_relation}) into the Kerr accretion rate formula, Eq.~(\ref{eq_kerr_mdot}). This gives the relativistic accretion rate of a Kerr PBH in an expanding universe as
\begin{align}
\label{eq:mdot_cosmo_final}
\dot{M} = 2\pi \rho_{\infty} \int_0^{\pi} & \left[1 - v(r)^2\right]^{\frac{1}{2\omega}} \left[ \mathcal{F}(r, \theta, a_*) \right]^{-\frac{1}{2\omega}} \frac{v(r)}{\sqrt{1-v(r)^2}} \sqrt{\Delta\Sigma} \sin\theta \, d\theta  .
\end{align}
This expression encapsulates the full dependence of the accretion rate on the PBH mass $M$ and spin $a_*$ (via the geometric terms $\mathcal{F}$, $\Delta$, and $\Sigma$), as well as on the cosmological epoch through the background density $\rho_{\infty}(t)$ and the Hubble parameter $H(t)$ (which is present in $v(r)$).

\subsection{The Critical Accretion Rate}

\begin{figure}[t]
\centering
\includegraphics[width=.49\textwidth]{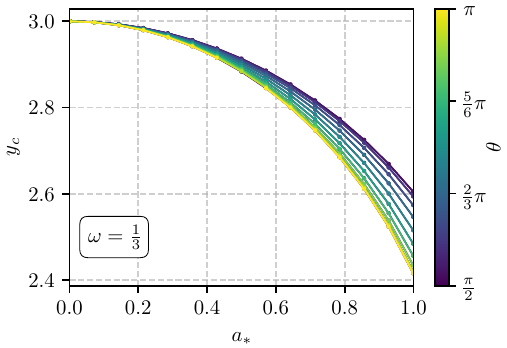}
\hspace{-0.1cm}
\includegraphics[width=.49\textwidth]{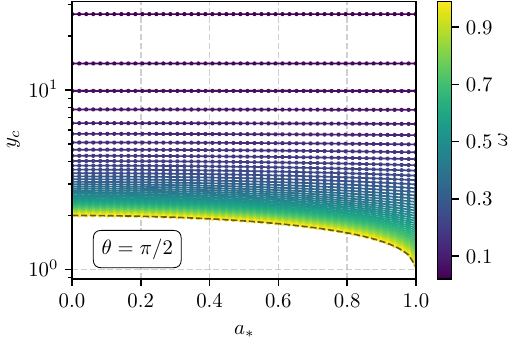}
\caption{\textbf{Left Panel:} Plot of the critical point $y_c \equiv (a x / GM)_c$ as a function of the spin parameter $a_*$ for different polar angles $\theta$, indicated by the color bar, with the equation of state $\omega = 1/3$. \textbf{Right Panel:} Same as the left panel, but for the equatorial plane ($\theta = \pi/2$) and various values of the equation of state $\omega$, shown by the color bar. The black dashed line corresponds to the location of horizon $r_+$.}
\label{Two_plots_a}
\end{figure}

\begin{figure}[t]
\centering
\includegraphics[width=.499\textwidth]{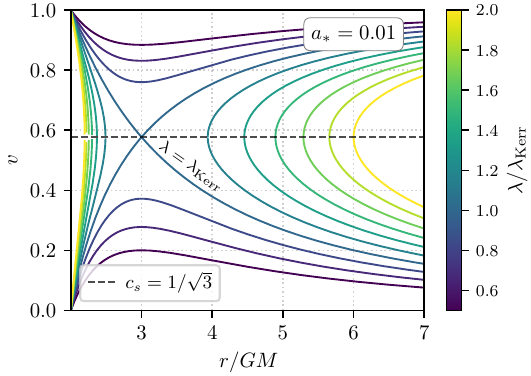}
\hspace{-0.2cm}
\includegraphics[width=.499\textwidth]{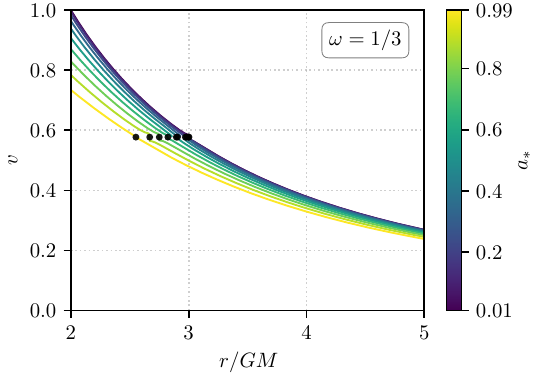}
\caption{\textbf{Left Panel:} The radial velocity $v(r)$ as a function of dimensionless radius $r/GM$ for a fixed spin parameter $a_* = 0.01$ and equation of state $\omega = 1/3$. The color bar visualizes the variation in the effective accretion rate parameter, $\lambda / \lambda_{\rm Kerr}$, showing how the subsonic and supersonic branches evolve. The flow becomes critical only when the accretion rate equals the critical value, $\lambda = \lambda_{\rm Kerr}$, where the subsonic and supersonic roots merge at the sonic point, marked by the critical speed $c_s = 1/\sqrt{3}$. \textbf{Right Panel:} The flow velocity $v$ at the critical accretion rate is shown as a function of $r$ for an equation of state $\omega = 1/3$. The colour bar indicates the corresponding values of the spin parameter $a_*$. The black dots mark the associated critical points.}
\label{fig:v_profile}
\end{figure}

The equations derived so far describe a family of possible accretion solutions. To find the unique, physically realized accretion rate, we must find the critical point of the flow, where the flow transitions from subsonic to supersonic (see the left panel of Fig.~\ref{fig:v_profile})~\cite{Dihingia:2018tlr}. The critical point can also be realized as the maximum accretion rate that the system can sustain. We find this point by maximizing the mass flux with respect to both the peculiar velocity, $v_p$ and the radial distance, $r$ ~\cite{Bondi:1952ni}.

First, we maximize the accretion rate from Eq.~(\ref{eq:mdot_cosmo_final}) with respect to the peculiar velocity $d\dot{M}/dv_p = 0$ at a fixed radius. The term containing the velocity dependence is $(1-v^2)^{(1-\omega)/2\omega} v$. Setting the derivative of this term to zero gives the velocity at the critical point
\begin{equation}
    v_p^c = \sqrt{\omega}-Hax .
\end{equation}
This means the peculiar velocity at the critical point is the speed of sound, $\omega = p/\rho$, of the fluid minus the Hubble flow, as expected. Next, we insert this critical velocity back into the expression for $\dot{M}$ and maximize the rate with respect to the comoving radial distance, $y \equiv ax/GM$. This procedure $d\dot{M}/dy = 0$ yields a complex polynomial equation as
\begin{align}
& \;\; 2\omega y^7 - y^6(1 + 3\omega)+ y^5\left(3a_*^2\omega + 3a_*^2\omega\cos^2\theta\right)  + y^4\left(-2a_*^2 + a_*^2\omega + a_*^2\cos^2\theta - 8a_*^2\omega\cos^2\theta\right) \notag \\
& + y^3\left(4a_*^2 - 6a_*^2\omega + a_*^4\omega - 4a_*^2\cos^2\theta + 6a_*^2\omega\cos^2\theta + 4a_*^4\omega\cos^2\theta + a_*^4\omega\cos^4\theta\right)  \notag \\
& + y^2\left(-a_*^4 + 2a_*^4\omega + 2a_*^4\cos^2\theta - 4a_*^4\omega\cos^2\theta - 3a_*^4\omega\cos^4\theta\right) \notag \\
& + y\left(-2a_*^4\omega\cos^2\theta + a_*^6\omega\cos^2\theta + 2a_*^4\omega\cos^4\theta + a_*^6\omega\cos^4\theta\right) + (a_*^6\cos^2\theta - a_*^6\omega\cos^4\theta) =0 .
\end{align}
Due to the spin of the BH, the potential is no longer spherically symmetric, and thus the solution of the polynomial, $y_c$ is a function of both the spin parameter $a_*$ and the polar angle $\theta$, i.e. $y_c = y_c(a_*, \theta)$. This equation must be solved numerically for each value of the spin parameter $a_*$ and polar angle $\theta$, as the critical point's location is not spherically symmetric. The results of this numerical solution are presented in Fig.~\ref{Two_plots_a}. The plot shows the location of the critical radius $y_c \equiv (ax/GM)_c$ as a function of the BH's spin parameter $a_*$. The most significant trend is that the critical radius $y_c$ decreases as the spin parameter $a_*$ increases. This means that for more rapidly rotating BHs, the fluid's sonic point moves closer to the event horizon. This is a direct consequence of relativistic frame-dragging, which enhances the effective gravitational pull on the accreting fluid, allowing it to become supersonic at a smaller radius.
Two key features are immediately apparent from the plot: 
\begin{itemize}
\item \textbf{Dependence on Polar Angle ($\theta$):} For any given non-zero spin, the critical radius is smallest at the poles and largest at the equator. The fluid accreting along the equatorial plane feels the strongest rotational effects and thus has its critical point goes far away from the BH.
\item \textbf{Dependence on EOS ($\omega$):} In the limit of a stiff fluid ($\omega \rightarrow 1$), the sound speed approaches the speed of light ($c_s = \sqrt{\omega} \rightarrow 1$). For accretion to occur, the fluid must become supersonic; however, this would require it to move faster than light, consequently, the sonic point shifts towards the event horizon $r_+$.
\end{itemize}

\begin{figure}[t]
\centering
\includegraphics[width=.495\textwidth]{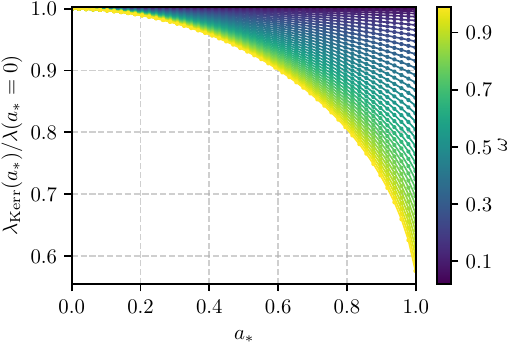}
\hspace{-0.2cm}
\includegraphics[width=.465\textwidth]{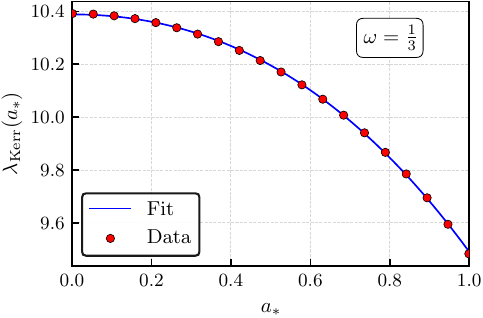}
\caption{\textbf{Left Panel:} Plot of the accretion efficiency parameter $\lambda_{\rm Kerr}(a_*)$ (Eq.~\ref{eq:lambda_kerr}) as a function of the Kerr spin parameter $a_*$ for different values of the equation of state $\omega$, indicated by the color bar. \textbf{Right Panel:} Variation of $\lambda_{\rm Kerr}(a_*)$ with the spin parameter $a_*$ for the equation of state $\omega = 1/3$. The red dots represent the numerical data, while the blue curve shows the best-fit result given by Eq.~\eqref{eq:lambda_fit}.}
\label{Two_plots_lamda_kerr}
\end{figure}

With the value of $y_c(a_*, \theta)$ determined, we can express the final critical accretion rate in a compact and conventional form as
\begin{equation}
\label{eq:mdot_final_parameterized}
    \dot{M} = 4\pi \lambda_{\text{Kerr}}(a_*) G^2 M^2 \rho_{\infty} .
\end{equation}
Here, $\lambda_{\text{Kerr}}(a_*)$ is the dimensionless \textit{accretion efficiency parameter} for the Kerr metric, which encapsulates all the complex physics of relativistic accretion in this geometry. It is defined by the integral
\begin{align}
\label{eq:lambda_kerr}
    \lambda_{\text{Kerr}}(a_*) = \frac{1}{2} \left(1-\omega\right)^{\frac{1-\omega}{2\omega}} \sqrt{\omega} \int_0^{\pi} & \left[ \mathcal{F}(y_c, \theta, a_*) \right]^{-\frac{1}{2\omega}}  \left\lbrace y_c^2 - 2y_c+a_{\ast}^2\right\rbrace^{\frac{1}{2}} \left\lbrace y_c^2+a^2_{\ast} \cos^2 \theta\right\rbrace^{\frac{1}{2}} \sin\theta \, d\theta ,
\end{align}
where the function $\mathcal{F}$ is evaluated at the critical radius $y_c(a_*, \theta)$. This parameter $\lambda_{\text{Kerr}}$ is the central result of our calculation, providing the correction factor to the standard Bondi accretion formula due to relativistic and rotational effects.

The numerical results are shown in the left panel of Fig.~\ref{Two_plots_lamda_kerr}. The plot displays the accretion efficiency $\lambda_{\text{Kerr}}(a_*)$ as a function of the spin parameter $a_*$. To isolate the effect of rotation, the efficiency is normalized to its non-rotating value, $\lambda(a_*=0) \simeq 10.4$. Our primary application is for PBHs in the early, radiation-dominated universe, where the equation of state is $\omega = 1/3$. To make the final evolution equations computationally tractable, we have performed a numerical fit to our calculated $\lambda_{\text{Kerr}}(a_*)$ values for this specific case (see the right panel of Fig.~\ref{Two_plots_lamda_kerr}). This provides a simple, analytical formula for the accretion efficiency as
\begin{equation}
\label{eq:lambda_fit}
\lambda_{\text{Kerr}}(a_*,\omega=1/3)  \simeq -a_{\ast}^2(0.255 \, a_{\ast}+0.645)+10.389 \; .
\end{equation}
This approximation will be used in all subsequent numerical and analytical calculations. Important to note that the effect of the spin parameter on $\lambda_{\text{Kerr}}$ is mostly dominated by its Schwzschild value. Therefore, while calculating the analytic expressions we assume the effect of spin to be negligible in the BH mass evolution.  
Nevertheless, from the plot, we can identify two primary trends:
\begin{itemize}
\item  \textbf{Dependence on Spin ($a_*$):} For any fluid with pressure ($\omega > 0$), the accretion efficiency monotonically decreases as the BH's spin $a_*$ increases. This is a key finding: a BH's rotation suppresses the accretion rate compared to a non-rotating BH of the same mass.
\item \textbf{Dependence on Equation of State ($\omega$):} The magnitude of this suppression is critically dependent on the fluid's ``stiffness''.
\begin{itemize}
\item In the stiff-fluid limit ($\omega \to 1$), the speed of sound $c_s$ approaches the speed of light, causing the critical radius $y_c$ to shift dramatically towards the event horizon $r_+$. This inward movement of the critical point effectively restricts the accretion volume, consequently leading to a significant decrease in the total mass accretion rate.
\item For a pressureless dust fluid ($\omega = 0$), the curve is perfectly flat. This means the accretion rate for pressureless matter is entirely independent of the BH's spin.
\end{itemize}
\end{itemize}
For a radiation-dominated fluid ($\omega = 1/3$), at maximal spin ($a_*=1$), the accretion rate is reduced to $\sim 91\%$ of its Schwarzschild value. This is the most relevant case for our study of PBHs in the early universe.

\begin{figure}[t]
\centering
\includegraphics[width=.499\textwidth]{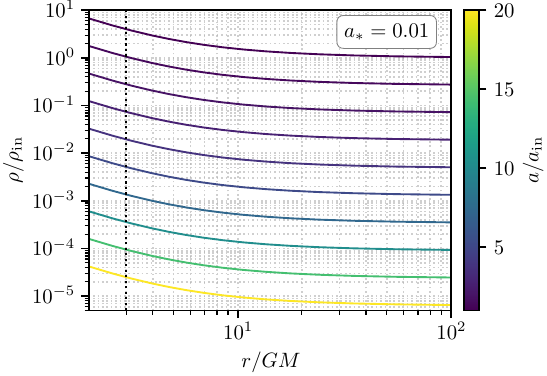}
\hspace{-0.2cm}
\includegraphics[width=.499\textwidth]{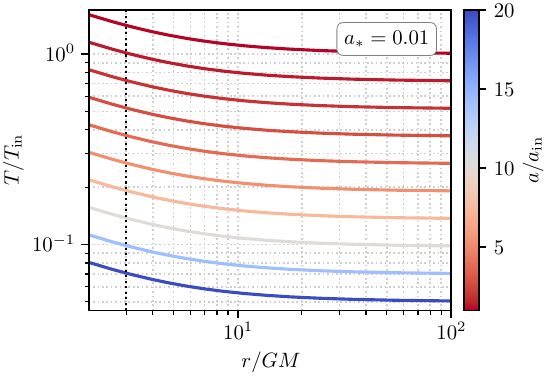}
\caption{The time evolution of the fluid's thermal properties is shown for a fixed PBH spin $a_* = 0.01$, calculated by substituting the critical transonic velocity profile into the fluid equations. The progression of time is shown by the color bar, representing the ratio of the instantaneous scale factor to the initial scale factor ($a/a_{\text{in}}$). \textbf{Left Panel:} The energy density profile, $\rho / \rho_{\text{in}}$, normalized by its initial value. The vertical dotted line marks the location of the critical radius ($y_c$). \textbf{Right Panel:} The corresponding fluid temperature profile, $T / T_{\text{in}}$, derived from the density profile using $\rho \propto T^4$. The decreasing temperature with the accretion time reflects the cosmological cooling of the background fluid.}
\label{fig:profile}
\end{figure}

In order to determine the radial velocity structure of the flow, we adopt the angle-averaging approximation, $v(r,\theta) \simeq v(r)$, and proceed by equating the general relativistic mass accretion rate \eqref{eq:mdot_cosmo_final} with the critical accretion rate $\dot{M}_{\text{crit}}$ parametrized by $\lambda_{\rm Kerr}$ [Eq.~\eqref{eq:mdot_final_parameterized}]. For an assumed equation of state parameter $\omega = 1/3$, this procedure results in the cubic equation $v^3(r) - v(r) + \mathcal{C}(r) = 0$, where the coefficient $\mathcal{C}(r)$ is defined as
\begin{equation}
    \mathcal{C}(r) = \frac{2\lambda_{\rm Kerr} G^2 M^2}{\mathcal{I}(r)}, \quad \text{with} \quad \mathcal{I}(r) = \int_0^{\pi} \mathcal{F}^{-\frac{3}{2}}\sqrt{\Delta \Sigma} \sin\theta d\theta.
\end{equation}
The left panel of Fig.~\ref{fig:v_profile} shows the velocity branches obtained by numerically solving this cubic equation for a fixed spin $a_* = 0.01$ and for various values of the rate parameter $\lambda/\lambda_{\rm Kerr}$, as indicated by the color bar. The physical solution is the unique ``transonic profile'' (represented by the critical curve $\lambda = \lambda_{\rm Kerr}$) that transitions smoothly from subsonic to supersonic velocity ($v = c_s = 1/\sqrt{3}$) at the critical radius $y_c$. In the right panel of Fig.~\ref{fig:v_profile}, we explicitly trace this physical transonic velocity profile, $v(r)$, at the critical accretion rate ($\lambda = \lambda_{\rm Kerr}$) for different PBH spins, where the variation in $a_*$ is shown by the color bar.

The derived radial velocity profile, $v(r)$, is subsequently inserted into the fluid continuity equation \eqref{eq:density_profile} to solve for the energy density profile $\rho(r)$. Since the background energy density $\rho_{\infty}$ is a function of cosmological time (or the scale factor $a$), the resulting density profiles are inherently time-evolving. The evolution of these density profiles is depicted in the left panel of Fig.~\ref{fig:profile}, where the color bar maps the progression of time via the normalized scale factor $a/a_{\text{in}}$. Furthermore, the corresponding temperature profile $T(r)$ is calculated directly from the density, assuming the fluid remains in thermal equilibrium: $\rho = (\pi^2/30) g_* T^4$. Using the effective number of relativistic degrees of freedom $g_* \simeq 106.75$, the time-evolving temperature profiles are presented in the right panel of Fig.~\ref{fig:profile}, with the time evolution again visualized by the color bar.

The sharp radial gradient in these profiles confirms that efficient accretion can also lead  the formation of localized ``hotspots'' around the PBH, as opposed to the conventional mechanism from their Hawking radiation \cite{He:2022wwy, Hamaide:2023ayu}. Interplay between the accretion and evaporation in the formation of hotspot could be an interesting extension of our work, and may have potential impact on the thermal history of the universe and contributing to dark matter constraints~\cite{Gunn:2024xaq}. We take up this in our forthcoming publication. 
This accretion induced local thermal energy injection can also have significant impact on the Hawking spectrum in cosmological setting ~\cite{Kalita:2025foa}.

\subsection{Accretion in a Cosmological Background}

\begin{figure}[t]
\centering
\includegraphics[width=.59\textwidth]{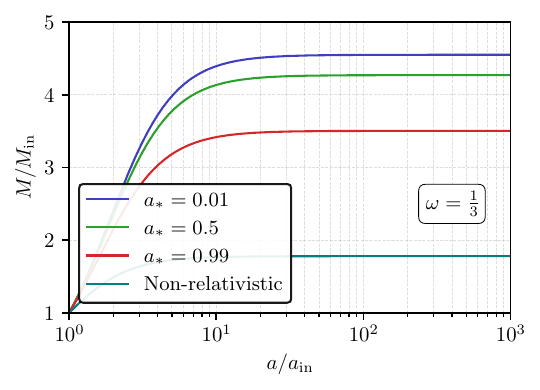}
\caption{The evolution of the PBH mass ratio $M/M_{\rm in}$ as a function of the cosmic scale factor $a/a_{\rm in}$ for a radiation-dominated universe ($\omega=1/3$). The different curves show the mass growth for fixed spin parameters: $a_* = 0.01$ (blue), $a_* = 0.5$ (green), and $a_* = 0.99$ (red). The non-relativistic case (cyan) is plotted for comparison, demonstrating that relativistic effects significantly enhance accretion, while spin suppresses it.}
\label{Two_plots_b}
\end{figure}

With the critical accretion rate $\dot{M}$ (Eq.~\ref{eq:mdot_final_parameterized}) parameterized by $\lambda_{\text{Kerr}}(a_*)$, we can now determine the evolution of the PBH mass. To solve it, we model the background cosmology as a flat Friedmann–Lema\^{i}tre–Robertson–Walker (FLRW) universe dominated by a fluid with a general equation of state $p = \omega \rho$. For our application to the early radiation-dominated era, $\omega = 1/3$. The background energy density $\rho_{\infty}$ and Hubble parameter $H$ evolve as
\begin{equation}
\rho_{\infty}(t) =3 M_p^2 H(t)^2 , \quad \text{and} \quad H(t) = \frac{2}{3(1+\omega)t} ,
\end{equation}
where $M_p = 1/\sqrt{8 \pi G} \simeq 2.435 \times 10^{18}\, \rm Gev$ is the reduced Planck mass. Substituting these into Eq.~(\ref{eq:mdot_final_parameterized}), we integrate the equation from the PBH formation time, $t_{\rm in}$, where the mass is $M_{\rm in}$, to a later time $t$ to give
\begin{equation}
\label{eq:mass_evolution}
\frac{M(t)}{M_{\rm in}} = \left[ 1-\frac{\lambda_{\text{Kerr}}(a_*) \gamma}{2(1+\omega)}\left( \frac{\frac{3}{2}H_{\rm in}(1+\omega)(t-t_{\rm in})}{\frac{3}{2}H_{\rm in}(1+\omega)(t-t_{\rm in})+1} \right) \right]^{-1} .
\end{equation}
The initial mass $M_{\rm in}$ is assumed to be a fraction $\gamma$ of the particle horizon mass at formation, given by $M_{\rm in} = 4 \pi \gamma M_p^2 H_{\rm in}^{-1}$, where $\gamma \sim 0.2$ is the collapse efficiency~\cite{Carr:1974} and $t_{\rm in} = 2 / (3(1+\omega)H_{\rm in})$. This equation describes the complete growth of the PBH mass due to relativistic accretion shown in Fig.~\ref{Two_plots_b} in the radiation dominated Universe $(\omega =1/3)$. Further note that in the above approximate derivation of PBH mass, we assumed spin evolution having insignificant effect, which is also consistent with our numerical derivation. 

A crucial quantity for cosmology is the final, asymptotic mass $M_{\rm acc}$ that the PBH achieves after the accretion epoch has effectively ended (i.e., as $t \to \infty$). In this limit, the term in the large parentheses in Eq.~(\ref{eq:mass_evolution}) approaches unity. This gives the final accreted mass 
\begin{equation}
\label{eq:mass_final_acc}
M_{\rm acc} \simeq M_{\rm in} \left( 1 - \frac{\lambda_{\text{Kerr}}(a_*) \gamma}{2(1+\omega)} \right)^{-1} .
\end{equation}
This is a key analytical result of our work. It directly links the final mass of the PBH to its spin $a_*$ (through our calculated $\lambda_{\text{Kerr}}$ parameter) and the background cosmology $\omega$. In a radiation-dominated era, for instance, the mass growth decreases with increasing spin: for $a_* = 0.01$ we obtain $M_{\rm acc} \simeq 4.53 M_{\rm in}$, for $a_* = 0.5$ it reduces to $M_{\rm acc} \simeq 4.25 M_{\rm in}$, and for the near-extremal case $a_* = 0.99$ it becomes $M_{\rm acc} \simeq 3.45 M_{\rm in}$, as illustrated in Fig.~\ref{Two_plots_b}. We therefore, found close to $40-50 \%$ increase of PBH mass due to relativistic accretion.

It is important to note that our baseline $40-50\%$ mass increase assumes a monochromatic formation fraction $\gamma \simeq 0.2$. In reality, PBH formation is governed by critical collapse~\cite{Musco:2023dak, Musco:2004ak}, meaning $\gamma$ follows a distribution rather than a single constant value. Because the accretion rate scales as $\dot{M} \propto M^2$, Eq.~(\ref{eq:mass_final_acc}) reveals a non-linear dependence on $\gamma$, characterized by a critical runaway threshold, $\gamma_c = 2(1+\omega)/\lambda_{\text{Kerr}}(a_*)$. If a PBH forms with an initial mass fraction approaching the critical value $\gamma \rightarrow \gamma_c$, its accretion rate outpaces the background cosmological dilution, leading to runaway accretion.
Consequently, integrating over the full critical collapse mass spectrum would skew the population-averaged mass growth $\langle M_{\rm acc} \rangle$ significantly higher than our monochromatic estimate, driven disproportionately by the heavy PBHs in the $\gamma \to \gamma_c$ tail.

It is crucial to reconcile the existence of this mathematical runaway threshold, $\gamma_c$, with the seminal theorem of Carr \& Hawking~\cite{Carr:1974nx}, which established that PBHs cannot undergo runaway accretion in a radiation-dominated universe. Our findings are entirely consistent with their fundamental conclusion. For typical PBH formation scenarios ($\gamma \simeq 0.2$), the condition $\gamma \lesssim \gamma_c$ is  satisfied. In this regime, the rapid cosmological dilution of the radiation fluid successfully chokes off the infall, resulting in a bounded mass increase rather than runaway growth. The runaway limit, $\gamma \rightarrow \gamma_c$, represents an extreme scenario that can only be realized if the initial density perturbation is anomalously large.

The characteristic timescale of accretion can be estimated by assuming that the PBH has effectively reached its final accreted mass $M_{\rm acc}$ at the moment when the accretion rate becomes comparable to the Hubble expansion rate (i.e., when $\dot{M} \simeq M H$). Under this approximation, the timescale is
\begin{equation}
\tau_{\rm BH}^{\rm acc} = t_{\rm acc} - t_{\rm in} \simeq \left( \frac{6\gamma\lambda_{\rm Kerr}(a_*)}{8-3\gamma\lambda_{\rm Kerr}(a_*)}-1 \right)\frac{M_{\rm in}}{8\pi\gamma M_p^2} .
\end{equation}
As an illustrative example, for a PBH with initial spin $a_* = 0.01$ and mass $10\,\mathrm{g}$ accreting in a radiation-dominated background, accretion effectively shuts off at approximately $\tau_{\rm BH}^{\rm acc} \simeq 3 \times 10^{6}\, M_p^{-1}$.

\section{PBH Mass Evolution: Hawking Evaporation}\label{sec:mass_evp}

To obtain the complete evolution of the PBH mass, we must now incorporate the mass-loss rate due to Hawking evaporation. The emission spectrum of Hawking radiation from a Kerr BH depends on the BH's rotation, the spin of the emitted particles, and the corresponding greybody factors that encode the backscattering of radiation by the curved spacetime geometry. The differential particle emission rate per unit momentum for the $i^{\text{th}}$ particle species with spin $s_i$, taking into account the particle mass, is given by~\cite{Hawking:1975vcx}
\begin{equation}
\frac{d^2 N_i}{dp\,dt}
= \frac{g_i}{2\pi} 
\sum_{l=s_i}^{\infty} \sum_{m=-l}^{l}
\frac{\Gamma_{s_i}^{lm}}
{\exp\!\left(\frac{E - m\Omega}{T_{\text{BH}}}\right)
- (-1)^{2s_i}}
\frac{p}{E} ,
\end{equation}
where the factor $(-1)^{2s_i}$ distinguishes bosons ($s_i$ integer) from fermions ($s_i$ half-integer) and $g_i$ is the number of internal degrees of freedom of the species $i$. Here $T_{\rm BH}$ and $\Omega$ are the Hawking temperature and horizon angular velocity for a Kerr BH
\begin{align}
T_{BH} &= \frac{1}{4 \pi G M } \frac{\sqrt{1-a^2_{\ast}}}{1+\sqrt{1-a^2_{\ast}}}, \\
\Omega &= \frac{a_{\ast}}{2 GM}\frac{1}{1+\sqrt{1-a^2_{\ast}}}.
\end{align}
The greybody factor $\Gamma_{s_i}^{lm}$ is related to the absorption cross section $\sigma_{s_i}^{lm}$ via
\(\Gamma_{s_i}^{lm} = \sigma_{s_i}^{lm} p^2 / \pi \), 
which allows us to write
\begin{equation}
\frac{d^2 N_i}{dp\,dt} =
\frac{g_i}{2\pi^2}
\sum_{l=s_i}^{\infty} \sum_{m=-l}^{l}
\frac{\sigma_{s_i}^{lm}}
{\exp\!\left(\frac{E - m\Omega}{T_{\text{BH}}}\right)
- (-1)^{2s_i}}
\frac{p^3}{E}.
\end{equation}
In the geometrical optics (eikonal) limit, the absorption cross section approaches the classical capture cross section for null geodesics, 
$\sigma_{s_i}^{lm} \to 27\pi G^2 M^2$. It is then convenient to define a dimensionless absorption coefficient $\varphi_{s_i}^{lm} = \sigma_{s_i}^{lm} / (27\pi G^2 M^2)$, so that the emission rate becomes
\begin{equation}
\frac{d^2 N_i}{dp\,dt} 
= g_i \sum_{l=s_i}^{\infty} \sum_{m=-l}^{l}
\frac{d^2 N_i^{lm}}{dp\,dt},
\end{equation}
where the partial contribution from each $(l,m)$ mode is given by
\begin{equation}
\frac{d^2 N_i^{lm}}{dp\,dt} =
\frac{1}{2\pi^2}
\sum_{l=s_i}^{\infty} \sum_{m=-l}^{l}
\frac{27\pi G^2 M^2\,\varphi_{s_i}^{lm}}
{\exp\!\left(\frac{E - m\Omega}{T_{\text{BH}}}\right)
- (-1)^{2s_i}}
\frac{p^3}{E}.
\end{equation}

\begin{figure}[t]
\centering
\includegraphics[width=.49\textwidth]{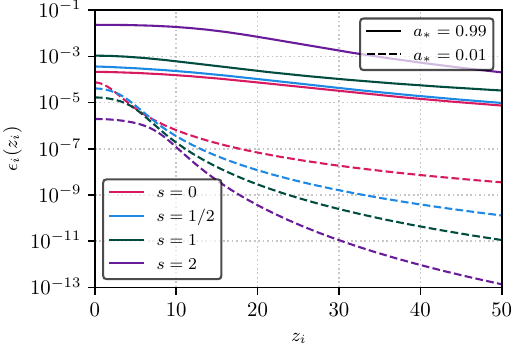}
\hspace{-0.3cm}
\includegraphics[width=.499\textwidth]{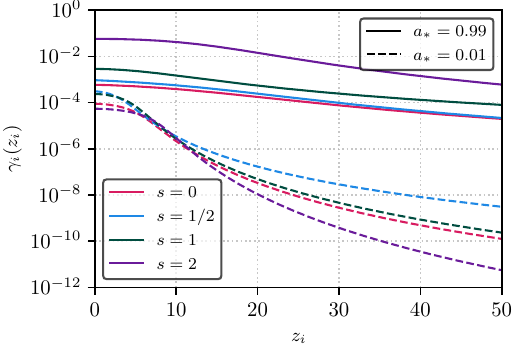}
\caption{The integrated particle emission efficiency parameter  ($\epsilon_i$) 
(\textbf{Left Panel})  and the corresponding spin-down efficiency parameter ($\gamma_i$) (\textbf{Right Panel}) are plotted as functions of the dimensionless particle mass parameter $z_i = 8\pi GM\mu_i$. Results are shown for various particle spins ($s = 0, 1/2, 1, 2$) and two distinct PBH spin scenarios: near-maximal Kerr ($a_* = 0.99$, solid lines) and Schwarzschild ($a_* = 0.01$, dashed lines). The figure demonstrates the strong dependence of emission efficiency on both the particle spin $s$ and the PBH spin $a_*$.}
\label{fig:ep_gam}
\end{figure}

\subsection{Mass Decay Rate}

The total mass loss rate is found by integrating the energy $E$ of all emitted particles over the distribution, summing over all species and angular modes ($l, m$) as~\cite{Page:1976df}
\begin{equation}\label{eq:evap_final_form}
\frac{dM}{dt} = -\sum_i \int_0^\infty E_i \frac{d^2 N_i}{dp dt} dp = -\epsilon \frac{M_p^4}{M^2},
\end{equation}
where $\epsilon$ is the dimensionless \textit{total evaporation efficiency}, given by the sum over all active particle species $\epsilon = \sum_i g_i \epsilon_i$, and the efficiency for a single species $\epsilon_i$ is given by the dimensionless integral
\begin{equation}
\epsilon_i = \frac{27}{128\pi^3} \int_{z_i}^{\infty} \sum_{l=s_i} \sum_{m=-l}^{l} \frac{\varphi_{s_i}^{lm}}{e^{\frac{x-m\Omega'}{2 f(a_{\ast})}}- (-1)^{2 s_i}} (x^2-z_i^2) x \, dx .
\end{equation}
This is expressed in terms of the dimensionless quantities $x = 8\pi GME$, $z_i = 8\pi GM\mu_i$, $\Omega' = 8\pi GM\Omega$, $f(a_*) = (4\pi G M) T_{BH}$ and $\mu_i$ is the mass of the emitted particle. It is crucial to note that the efficiency $\epsilon$ is not a constant. It explicitly depends on the spin $a_*$ and on the mass $M$ itself. 
As the PBH evaporates and $M$ decreases, its temperature $T_{BH} \sim 1/M$ rises, allowing more massive particles to be emitted and increasing the value of $\epsilon$.

In the Schwarzschild limit ($a_* \rightarrow 0$) and also in geometric–optics approximation, the epsilon parameter simplifies to $\epsilon \simeq \frac{27}{4} \frac{g_{\ast}(T_{\rm BH}) \pi}{480}$ with $g_{\ast}(T_{\rm BH}) \simeq 106.76$ denoting the number of relativistic degrees of freedom at the BH temperature $T_{\rm BH}=M_p^2/M$. Under this approximation, the PBH mass evolves as
\begin{equation}
M(t) \simeq 
M_{\rm in} \left( 1 - \frac{\lambda_{\text{Kerr}}(a_*) \gamma}{2(1+\omega)} \right)^{-1} \left[1 - \frac{t - t_{\rm acc}}{\tau_{\rm BH}^{\rm ev}} \right]^{1/3},
\end{equation}
where the evaporation timescale defined by $M \rightarrow 0$ is
\begin{eqnarray} \label{pbh-life}
 \tau_{\rm BH}^{\rm ev} \simeq \frac{M_{\rm in}^3}{3 \epsilon M_p^4} \left( 1 - \frac{\lambda_{\text{Kerr}}(a_*) \gamma}{2(1+\omega)} \right)^{-3}.
\end{eqnarray}
For instance, a PBH of mass $10\,\mathrm{g}$ formed in a radiation era evaporates at approximately $\tau_{\rm BH}^{\rm ev} \simeq 8.8 \times 10^{17}\, M_p^{-1}$. This demonstrates the large hierarchy between accretion and evaporation timescales with $\tau_{\rm BH}^{\rm ev} \gg \tau_{\rm BH}^{\rm acc}$, which will be useful for our analytical calculation in two different limits.

\begin{figure}[t]
    \centering
    \includegraphics[width=0.5\linewidth]{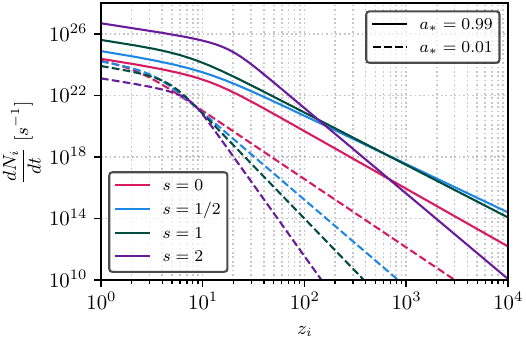}
    \caption{The particle production rate $dN_i/dt$ [Eq.~\eqref{eq:dndt_final}] is plotted as a function of the dimensionless particle mass $z_i = 8\pi GM\mu_i$. The curves show the strong dependence on both the particle spin $s$ and the PBH spin $a_*$ (solid for near-maximal Kerr, dashed for Schwarzschild). 
    }
    \label{fig:psi}
\end{figure}

\subsection{Angular Momentum Evolution}

As the BH evaporates, the emitted particles carry away angular momentum, causing the BH to spin down. The total rate of angular momentum loss, $dJ/dt$, is found by integrating the particle emission rate, weighted by the azimuthal quantum number $m$ as~\cite{Auffinger:2022khh, Arbey:2019jmj, Ewasiuk:2025dwn}
\begin{equation}\label{eq:djdt_final}
\frac{dJ}{dt} = -\sum_i \int_0^{\infty} \sum_{l=s_i} \sum_{m=-l}^{l} m \frac{d^2 N_i^{lm}}{dp dt} dp = -a_{\ast} \gamma  \frac{M_p^2}{M},
\end{equation}
where $\gamma = \sum_i \gamma_i$ is the dimensionless \textit{total spin-down efficiency}. The efficiency per species, $\gamma_i$, is defined by the integral
\begin{equation}
\gamma_i = \frac{27}{128\pi^3} \int_{z_i}^{\infty} \sum_{l=s_i} \sum_{m=-l}^{l} \frac{(m/a_*) \varphi_{s_i}^{lm}}{e^{\frac{x-m\Omega'}{2 f(a_{\ast})}}- (-1)^{2 s_i}} (x^2-z_i^2) \, dx .
\end{equation}
To find the evolution of the dimensionless spin parameter $a_*$ itself, we use the chain rule with the definition $J = a_* M^2 / (8\pi M_p^2)$. Combining this with our previously derived expressions for $dM/dt$ (Eq.~\ref{eq:evap_final_form}) and $dJ/dt$ (Eq.~\ref{eq:djdt_final}), we find the final evolution equation for the spin parameter
\begin{equation}
\label{eq:dadt_final}
\frac{da_*}{dt} = - a_* \left( 8\pi\gamma(M, a_*) - 2\epsilon(M, a_*) \right) \frac{M_p^4}{M^3}.
\end{equation}
This equation shows that the spin parameter evolves based on a competition between the angular momentum loss (proportional to $\gamma$) and the mass loss (proportional to $\epsilon$).

To accurately determine the BH decay and spin-down effect during PBH evaporation, especially in the near-extremal Kerr regime, we must compute the evaporation efficiency and spin-down efficiency parameters, $\epsilon_i$ and $\gamma_i$, respectively. These integrals replace the simplified constant greybody factors by incorporating the full frequency and angular dependence of the Kerr metric's greybody factor ($\varphi_{s_i}^{lm}$). We numerically evaluate these integrals for various particle spins ($s=0, 1/2, 1, 2$) and plot them in Fig.~\ref{fig:ep_gam}. The resulting figure clearly illustrates the strong dependence of the emission efficiency on both the particle spin and the BH spin. The emission efficiency generally decreases with increasing $s$ (due to the higher gravitational barrier) and increases with increasing $a_*$ (due to frame-dragging effects).

\subsection{Total Particle Production}

The total particle production rate for a species $i$, $\Gamma_{BH \to i}$, is found by integrating the differential particle number flux over all momenta as
\begin{align}\label{eq:dndt_final}
\Gamma_{BH \to i} &= \frac{dN_i}{dt} = \int_0^{\infty} \frac{d^2 N_i}{dp dt} \, dp = g_i \, \psi_i(M, a_*) \frac{M_p^2}{M},
\end{align}
where $\psi_i$ is the dimensionless \textit{particle number efficiency} given by
\begin{equation}
\psi_i = \frac{27}{128\pi^3} \int_{z_i}^{\infty} \sum_{l=s_i} \sum_{m=-l}^{l} \frac{\varphi_{s_i}^{lm}}{e^{\frac{x-m\Omega'}{2 f(a_{\ast})}}- (-1)^{2 s_i}} (x^2-z_i^2) \, dx.
\end{equation}
Like $\epsilon$ and $\gamma$, the function $\psi_i$ depends on the PBH mass $M$ (through the integration limit $z_i = 8\pi G M \mu_i$) and the spin $a_*$, and must be calculated numerically. The parameter $\psi_i$ serves as the effective average of the greybody factor, $\varphi_{s_i}^{lm}$, weighted across all emission frequencies, angular momenta ($l$), and azimuthal modes ($m$). The overall trends of the particle production rate $dN_i/dt$, as illustrated in Fig.~\ref{fig:psi}, show that faster-spinning PBHs are significantly more efficient at emitting all particle species, particularly those with higher spin values ($s$).

Therefore the total number of particle produced by a BH, $N_i$, is the time-integral of the particle production rate (Eq.~\ref{eq:dndt_final}) over the BH's entire lifetime, from its formation ($a_{\text{in}}$) to its evaporation ($a_{\text{ev}}$) as
\begin{equation}\label{total_N_i}
N_{i} = \int_{t_{\text{in}}}^{t_{\text{ev}}} \Gamma_{\text{BH} \to i}(t) \, dt = \int_{t_{\text{in}}}^{t_{\text{ev}}} g_{i} \psi_{i}(M(t), a_*(t)) \frac{M_p^2}{M(t)} \, dt .
\end{equation}
This integral must be solved numerically, as the emission rate $\psi_{i}$ depends on the evolving mass $M(t)$ and spin $a_*(t)$.

To derive a concise, analytical estimate for the total particle production yield, we utilize two major physical simplifications relevant to our model. First, anticipating the results of the next section~\ref{sec:mass_acc_evp}, where we show that accretion eventually drives the PBH spin to zero, we are justified in adopting the non-spinning limit ($a_* \to 0$) for this calculation. Second, for the mass range of interest, the dimensionless PBH temperature term, $z_i = 8 \pi GM \mu_i$, is small ($z_i \to 0$). In this combined limit, the factor $\psi_i$, can be simplified into a spin-dependent constant term~\cite{Cheek:2021odj, Ukwatta:2015iba}
\begin{equation}\label{eq:psi_fit}
\psi_i \simeq \frac{27}{128\pi^3} \times 
\begin{cases}
2.45  & s=0 \quad (\text{scalar}) \\
0.897  & s=1/2 \quad (\text{fermion}) \\
0.273  & s=1 \quad (\text{vector}) \\
0.026 & s=2 \quad (\text{graviton})
\end{cases}.
\end{equation}
We then distinguish between two scenarios based on the relationship between the particle mass $\mu_i$ and the PBH temperature $T_{\rm BH}^{\rm acc} \simeq T_{\rm BH}^{\rm in} \left(1-\frac{3\lambda_{\rm Kerr(a_*)} \gamma}{8}  \right)$ at the end of the dominant accretion phase:

\noindent
\textbf{Case I: Light Particles ($\mu_i \lesssim T_{\rm BH}^{\rm acc}$)}

If the particle mass is below the PBH temperature ceiling attained after accretion ($T_{\rm BH}^{\rm acc}$), the particle production mechanism is effectively active throughout the entire PBH lifetime (from initial time $t_{\rm in}$ until evaporation time $t_{\rm ev}$). Integrating the particle production rate $\frac{dN_i}{dt}$ [Eq.~\eqref{eq:dndt_final}] over this full timescale and employing the simplified $\psi_i$ values from Eq.~\eqref{eq:psi_fit}, we obtain the total number yield
\begin{equation}
N_i^{\mu_i \lesssim T_{\rm BH}^{\rm acc}} = \int_{t_{\rm in}}^{t_{\rm ev}} \frac{dN_i}{dt} dt \simeq \frac{15 g_i}{2g_* \pi^4}\frac{M_{\rm acc}^2}{M_p^2} \times
\begin{cases}
2.45  & s=0 \\
0.897  & s=1/2 \\
0.273  & s=1 \\
0.026 & s=2
\end{cases}.
\end{equation}

\noindent
\textbf{Case II: Heavy Particles ($\mu_i > T_{\rm BH}^{\rm acc}$)}

The second distinct case arises for heavy particles whose mass exceeds the final accretion temperature ($\mu_i > T_{\rm BH}^{\rm acc}$). For these particles, emission is only kinematically possible when the PBH temperature falls below $\mu_i$. Thus, the integration starts at a delayed time $\tilde{t}_i$, defined as the time when the PBH mass (and corresponding temperature) satisfies the condition $\mu_i = T_{\rm BH}$. The expression for this onset time $\tilde{t}_i$ is given by
\begin{equation}
\tilde{t}_i = \tau_{\rm BH}^{\rm ev} \left(1- \frac{M_p^6}{\mu_i^3 M_{\rm acc}^3}  \right).
\end{equation}
Integrating the particle production rate only over the effective emission window from $\tilde{t}_i$ to the evaporation time $t_{\rm ev}$, we arrive at the yield for heavy particles
\begin{equation}
N_i^{\mu_i > T_{\rm BH}^{\rm acc}} = \int_{\tilde{t}_{i}}^{t_{\rm ev}} \frac{dN_i}{dt} dt \simeq \frac{15 g_i}{2g_* \pi^4}\frac{M_{p}^2}{\mu_i^2} \times
\begin{cases}
2.45  & s=0 \\
0.897  & s=1/2 \\
0.273  & s=1 \\
0.026 & s=2
\end{cases}.
\end{equation}


\section{PBH Final Mass and Spin Evolution}\label{sec:mass_acc_evp}

We can now assemble the final, coupled system of differential equations that govern the complete evolution of a spinning PBH. The net change in the PBH mass is the sum of the relativistic accretion rate (Eq.~\ref{eq:mdot_final_parameterized}) and the mass loss from Hawking evaporation (Eq.~\ref{eq:evap_final_form}).

\begin{figure}[t]
\centering
\includegraphics[width=.49\textwidth]{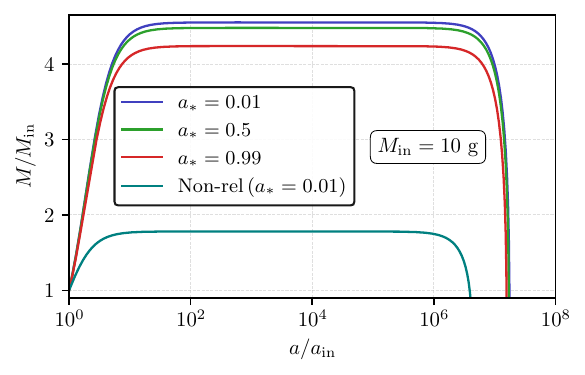}
\hspace{-0.3cm}
\includegraphics[width=.499\textwidth]{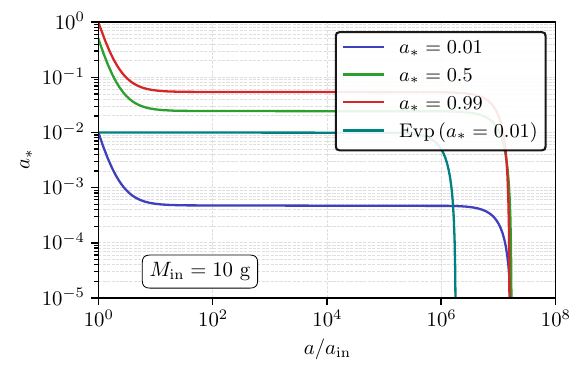}
\caption{\textbf{Left Panel:} The complete mass evolution ($M/M_{\text{in}}$) of a PBH with an initial mass of $M_{\text{in}} = 10 \text{g}$, plotted against the normalized cosmic scale factor $a/a_{\text{in}}$. The different curves correspond to different initial spin parameters: $a_* = 0.01$ (blue), $a_* = 0.5$ (green), and $a_* = 0.99$ (red). The purely non-relativistic accretion case (cyan, for $a_* = 0.01$) is shown for comparison. \textbf{Right Panel:} The evolution of the PBH spin parameter $a_*$ as a function of the cosmic scale factor $a/a_{\text{in}}$, for the same initial mass $M_{\text{in}} = 10 \text{g}$. The blue, green, and red curves show the complete evolution, including both relativistic accretion and Hawking evaporation, for initial spins of $a_* = 0.01, 0.5,$ and $0.99$, respectively, while, the cyan curve shows the `evaporation-only' case (starting at $a_*=0.01$).}
\label{total_mass_spin_evo}
\end{figure}

The final ordinary differential equations for the PBH mass $M$ is the gain from accretion minus the loss from evaporation as
\begin{equation}
\label{eq:final_dmdt}
\frac{dM}{dt} = \left( \frac{dM}{dt} \right)_{\text{Acc}} + \left( \frac{dM}{dt} \right)_{\text{Evap}} = 4\pi \lambda_{\text{Kerr}}(a_*) G^2 M^2 \rho_{\infty}(t) - \epsilon(M, a_*) \frac{M_p^4}{M^2} ,
\end{equation}
where $\lambda_{\text{Kerr}}(a_*)$ is given by the fit in Eq.~(\ref{eq:lambda_fit}) and $\rho_{\infty}(t)$ is the background energy density of the universe at time $t$. The spin evolution is derived from the total change in angular momentum, $J = a_* M^2 / (8\pi M_p^2)$. The total rate of change is $\frac{dJ}{dt} = (\frac{dJ}{dt})_{\text{Acc}} + (\frac{dJ}{dt})_{\text{Evap}}$. As per our accretion model ($u_\phi=0$), the accreting fluid carries no angular momentum, so $(\frac{dJ}{dt})_{\text{Acc}} = 0$. Therefore the final BH spin evolution will come only from evaporation given by Eq.~(\ref{eq:djdt_final}): $(\frac{dJ}{dt})_{\text{Evap}} = -a_* \gamma \frac{M_p^2}{M}$. Using the chain rule, $\frac{dJ}{dt} = \frac{1}{8\pi M_p^2} (M^2 \frac{da_*}{dt} + 2a_* M \frac{dM}{dt})$, and setting it equal to the total rate of change gives the final spin evolution equation as
\begin{equation}
\label{eq:final_dadt}
\frac{da_*}{dt} = -\frac{2a_*}{M} \left( \frac{dM}{dt} \right) - a_* \left( 8\pi\gamma(M, a_*) \right) \frac{M_p^4}{M^3} ,
\end{equation}
where $dM/dt$ is the \textit{total} mass rate of change from Eq.~(\ref{eq:final_dmdt}). This equation elegantly shows that the spin parameter $a_*$ changes for two reasons:
\begin{enumerate}
    \item \textbf{Dilution:} As the PBH accretes mass, its angular momentum is ``diluted'' over a larger mass, causing $a_*$ to decrease.
    \item \textbf{Evaporation:} The PBH preferentially sheds angular momentum via Hawking radiation, actively spinning it down.
\end{enumerate}
This coupled system of equations, Eq.~(\ref{eq:final_dmdt}) and Eq.~(\ref{eq:final_dadt}), forms the complete model for our cosmological analysis. 
The evolution of the PBH spin parameter $a_*$ is governed by the total change in mass $\dot{M}$. However, due to the substantial hierarchy between the timescales of mass loss via evaporation ($\tau_{\rm BH}^{\rm ev}$) and mass gain via accretion ($\tau_{\rm BH}^{\rm acc}$), we can simplify the system by initially considering only the accretion-driven spin evolution. Thus, we neglect the evaporation term and solve the relation $\frac{da_*}{dt} = -\frac{2a_*}{M} \dot{M}_{\rm acc}$. Integrating this simplified differential equation with the assumption that accretion dominates the mass change, the spin parameter $a_*^{\rm acc}$ after the mass saturates at $M_{\rm acc}$ is analytically approximated as
\begin{equation}
a_*^{\rm acc} \simeq a_*^{\rm in}\left( \frac{M_{\rm in}}{M_{\rm acc}} \right)^2.
\end{equation}
Substituting the expression for the maximum accreted mass, $M_{\rm acc}$ [Eq.~\eqref{eq:mass_final_acc}], we arrive at the final spin parameter following the accretion phase
\begin{equation}
a_*^{\rm acc} \simeq a_*^{\rm in}\left(1- \frac{\lambda_{\rm Kerr}(a_{*}^{\rm in})\gamma}{2(1+\omega)} \right)^2.
\end{equation}
This formula clearly demonstrates the de-spinning effect of accretion. For example, considering a radiation-dominated Universe ($\omega=1/3$), a highly spinning Kerr BH with $a_{*}^{\rm in} = 0.99$ reduces its spin significantly to $a_*^{\rm acc} \simeq 0.08$. Similarly, intermediate spins reduce from $a_{*}^{\rm in} = 0.5$ to $a_*^{\rm acc} \simeq 0.03$, and low spins from $a_{*}^{\rm in} = 0.01$ approach $a_*^{\rm acc} \simeq 5 \times 10^{-4}$, as further substantiated by the numerical results shown in the right panel of Fig.~\ref{total_mass_spin_evo}.

The total mass evolution ($M/M_{\text{in}}$) of a PBH with an initial mass of $M_{\text{in}} = 10 \text{g}$, is shown in the left panel of Fig.~\ref{total_mass_spin_evo} as a function of scale factor $a$. The plot illustrates the competition between relativistic accretion and Hawking evaporation. The different curves correspond to different initial spin parameters: $a_* = 0.01$ (blue), $a_* = 0.5$ (green), and $a_* = 0.99$ (red). The purely non-relativistic accretion case (cyan, for $a_* = 0.01$) is shown for comparison. The mass initially grows rapidly due to accretion, reaching a plateau. This mass gain is significantly enhanced by relativistic effects but is suppressed by higher spin. The PBH then maintains this mass until the evaporation phase, where it rapidly loses mass and disappears. The more massive, relativistically-accreted PBHs (with lower spin) survive for a longer cosmic time.

The evolution of the PBH spin parameter $a_*$ as a function of the cosmic scale factor $a/a_{\text{in}}$ is shown in the right panel of Fig.~\ref{total_mass_spin_evo} for the same initial mass $M_{\text{in}} = 10 \text{g}$. The blue, green, and red curves show the complete evolution, including both relativistic accretion and Hawking evaporation, for initial spins of $a_* = 0.01, 0.5,$ and $0.99$, respectively. The initial, rapid decrease in $a_*$ is a direct result of spin dilution: the accretion of zero-angular-momentum fluid increases the PBH's mass, which dilutes its specific angular momentum. After the accretion epoch ends, the spin parameter remains nearly constant. Finally, as the PBH begins its final evaporation (at $a/a_{\text{in}} \gtrsim 10^6$), it preferentially sheds angular momentum, causing $a_*$ to plummet to zero. The cyan curve shows the `evaporation-only' case (starting at $a_*=0.01$) for comparison, which lacks the initial spin-dilution phase.

Given that this initial phase of relativistic accretion proceeds extremely rapidly, over only a few cosmological e-foldd, the post-accretion mass $ M_{\rm acc} $ effectively acts as the initial mass of the PBH for practical purposes in its late-time evolution. Consequently, in the following sections, all observational constraints from relevant physical processes are evaluated using this semi-stationary mass $ M_{\rm acc} $ as the reference.

While the late-time phenomenology and observational bounds are directly sensitive only to this semi-stationary mass $M_{\rm acc}$, distinguishing it from the mass immediately following formation is theoretically crucial. Early universe models relate the primordial curvature power spectrum to an initial overdensity $\delta$, which in turn dictates the PBH mass via critical collapse, $M(\delta)$. The precise definition of $M(\delta)$---and the extent to which it implicitly incorporates immediate post-formation accretion---can vary depending on the specific modeling framework. However, standard calculations of $M(\delta)$ typically capture only the immediate, non-linear collapse dynamics. They generally omit the subsequent secular accretion over multiple e-folds and the strong, spin-dependent suppression of this mass growth. Therefore, our relativistic accretion framework serves as an essential physical transfer function. To satisfy a given phenomenological bound on $M_{\rm acc}$, the underlying inflationary model must map its predicted overdensities $\delta$ through this extended accretion phase to accurately bridge theoretical predictions with bottom-up observational constraints.

\section{The BBN Constraint}\label{sec:BBN}

We now apply our complete PBH evolution model to derive cosmological constraints. The first and most stringent constraint comes from BBN. High-energy particles injected into the cosmic plasma around the time of BBN ($T \sim 4 \text{ MeV}$~\cite{Hannestad:2004px}) can destroy the light elements (D, ${}^4$He, ${}^3$He, ${}^7$Li) after they have been synthesized, or alter the neutron-to-proton ratio before synthesis~\cite{Keith:2020jww, Boccia:2024nly, Wu:2025ovd}. This places a strong upper bound on the lifetime of an evaporating PBH. To be cosmologically safe, a PBH must evaporate \textit{before} BBN begins. We therefore impose the condition that the PBH's evaporation must be complete by the time the universe cools to $T_{\rm BBN} \approx 4 \text{ MeV}$. This allows us to set an upper limit on the \textit{initial} mass $M_{\rm in}$ of a PBH.

\begin{figure}[t]
\centering
\includegraphics[width=.49\textwidth]{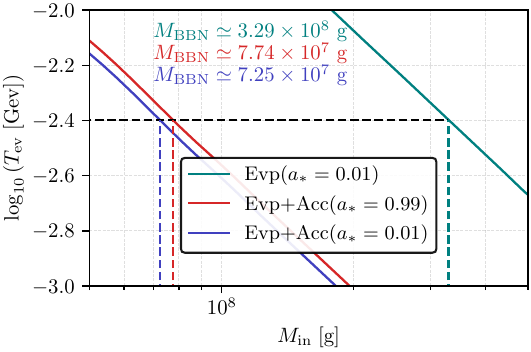}
\hspace{-0.1cm}
\includegraphics[width=.49\textwidth]{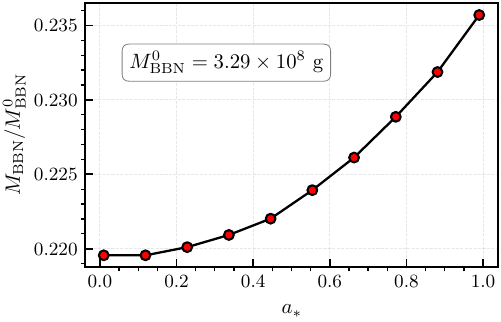}
\caption{\textbf{Left Panel:} The background temperature at the time of PBH evaporation ($T_{\text{ev}}$) as a function of the initial mass ($M_{\text{in}}$). The horizontal dashed line represents the BBN temperature limit. We compare the evaporation-only model (Evp, cyan curve) with our full model including accretion and evaporation (Evp+Acc) for high spin ($a_*=0.99$, red) and low spin ($a_*=0.01$, blue). Including accretion (blue and red curves) significantly lowers $T_{\text{ev}}$ for a given $M_{\text{in}}$, thus shifting the BBN mass limit (vertical dashed lines) to much smaller values. \textbf{Right Panel:} The BBN upper limit on $M_{\text{in}}$, $M_{\text{BBN}}$, as a function of the initial spin $a_*$. The mass limit is normalized to the evaporation-only case, $M_{\text{BBN}}^0 = 3.29 \times 10^8 \text{ g}$.}
\label{BBN_result}
\end{figure}

To find the constraint, we must relate the PBH's evaporation time to the background cosmic temperature. In a radiation-dominated universe, the scale factor $a$ and temperature $T$ are related by $a \propto 1/T$. We can therefore write $a/a_{\rm in} = T_{\rm in}/T$, where $a_{\rm in}$ and $T_{\rm in}$ are the scale factor and temperature at the PBH formation time. The initial temperature $T_{\rm in}$ is related to the initial PBH mass $M_{\rm in}$ as
\begin{equation}
    T_{\rm in} = M_{\rm in} (4 \pi \gamma)^{\frac{1}{2}} \left(\frac{\pi^2}{90}g_{*} \right)^{-\frac{1}{4}} \left(\frac{M_p}{M_{\rm in}} \right)^{\frac{3}{2}} ,
\end{equation}
where $g_* \simeq 106.75$ is the effective number of relativistic degrees of freedom~\cite{Husdal:2016haj}. 
Now the evaporation temperature ($T_{\rm ev}$) can be obtained by utilizing the expression of PBH lifetime (Eq. \eqref{pbh-life}) as,
\begin{equation}\label{24}
    \tau_{\rm BH}    
   \simeq\frac{1}{2H_{\rm acc}} \left(\frac{a_{\rm ev}}{a_{\rm acc}}\right)^{2} = \frac{\Min^3}{3 \epsilon M_p^4}\left( 1 - \frac{\lambda_{\text{Kerr}}(a_*^{\rm acc}) \gamma}{2(1+\omega)} \right)^{-3},
     \end{equation}
where, the Hubble parameter at the end of accretion $t_{\rm acc}$ is denoted as $H_{\rm acc}~(= 4\pi \gamma M_p^2 M_{\rm acc}^{-1})$. Now, in this radiation dominated Universe $w=1/3$, the energy density $\rho_r=\frac{\pi^2}{30} g_{\ast}(T) T^4$, and the scale factor and temperature are related as $a_{\rm ev}/a_{\rm acc}=T_{\rm acc}/T_{\rm ev}$. Therefore, the PBH evaporation temperature ($T_{\rm ev}$) is computed as,
\bea
T_{\rm ev} \simeq \sqrt{\frac{3\epsilon}{2}} \left( \frac{\pi^2 g_{\ast}}{30} \right)^{-\frac{1}{4}} \left( \frac{M_p}{\Min} \right)^{\frac{3}{2}} \left( 1 - \frac{3\lambda_{\text{Kerr}}(a_*^{\rm acc}) \gamma}{8} \right)^{\frac{3}{2}} M_p ,~~ \label{25}
\eea
where the relation \( T_{\rm acc} \simeq T_{\rm in} = (\pi^2 g_{\ast}/90)^{-1/4} (H_{\rm in} M_p)^{1/2} \) is used corresponding to radiation temperature at $H_{\rm acc}$.
Combining Eqs.~(\ref{24}) and (\ref{25}), the initial PBH mass can be expressed as a function of the evaporation temperature,
 Combining these relations, Eq.~(\ref{24}) and (\ref{25}), the initial PBH mass and its evaporation temperature can be related as, 
\begin{equation} \label{tevM}
    \left(\frac{\Min}{1\rm g} \right) \simeq 3.94\times10^8 
 \left( 1 - \frac{3\lambda_{\text{Kerr}}(a_*^{\rm acc}) \gamma}{8} \right) \left(\frac{T_{\rm ev}}{4 \rm Mev} \right)^{-2/3}.
\end{equation}
In the above expression, we consider $g_* = 106.75$ at the time of PBH formation. 
No accretion limit can be obtained by setting $\lambda_{\rm Kerr} =0$, which yields the well known BBN bound on PBH initial mass \( M_{\rm in} \simeq 3.94 \times 10^8 \, \mathrm{g} \).

    
By setting $T_{\rm ev} = T_{\rm BBN} \approx 4 \text{ MeV}$, we can numerically solve for the maximum allowed initial mass, $M_{\rm BBN}$, for different initial PBH spin $a_*$. A critical finding of our work is that \textit{relativistic accretion delays evaporation}. As shown in the left panel of Fig.~\ref{total_mass_spin_evo}, the PBH mass first grows significantly to $M_{\rm acc} > M_{\rm in}$. Since the evaporation timescale is highly sensitive to mass ($\tau_{\rm evap} \propto M^3$), this accreted mass causes the PBH to survive for a much longer cosmic time than predicted by models that ignore accretion.

Our numerical results, presented in Fig.~\ref{BBN_result}, confirm that relativistic accretion has a major impact on the BBN constraint. The left panel shows the cosmic temperature at the time of the PBH's final evaporation, $T_{\text{ev}}$, as a function of its initial mass $M_{\text{in}}$. The horizontal dashed black line at $T_{\text{BBN}} \approx 4 \text{ MeV}$ is the critical boundary. The cyan curve shows that in a standard, evaporation-only scenario, the BBN limit on the initial mass is $M_{\text{in}} \le 3.29 \times 10^8 \text{ g}$. However, when the relativistic accretion model is included, the PBH lifetime is significantly extended. For a given $M_{\text{in}}$, the PBH survives longer, and thus $T_{\text{ev}}$ is lower. This shifts the curves to the left. For a low-spin PBH ($a_*=0.01$, blue curve), the new constraint is $M_{\text{in}} \le 7.25 \times 10^7 \text{ g}$. This demonstrates that relativistic accretion strengthens the BBN constraint by a factor of $\sim 4.5$. The plot also reveals the crucial role of spin. The constraint for a high-spin PBH ($a_*=0.99$, red curve) is $M_{\text{in}} \le 7.74 \times 10^7 \text{ g}$. This is a weaker constraint than the low-spin case.

The right panel illustrates this spin dependence more clearly. It plots the final BBN mass limit $M_{\text{BBN}}$ (normalized to the evaporation-only limit $M_{\text{BBN}}^0$) as a function of $a_*$. The plot shows that $M_{\text{BBN}}$ monotonically increases with spin. This is the expected behavior: since higher spin suppresses the accretion efficiency $\lambda_{\text{Kerr}}$ (as shown in the right panel of Fig.~\ref{Two_plots_a}), the PBH's lifetime is not extended as much, thus relaxing the constraint.


\section{Dark Matter from Evaporating PBHs}\label{sec:DM}

In addition to BBN constraints, PBHs that evaporate can also \textit{produce} the DM observed today. If the DM is a stable particle that is emitted via Hawking radiation, the total abundance of this PBH-generated DM must not exceed the observed value, $\Omega_{\text{DM,0}}h^2 \approx 0.12$~\cite{Planck:2018vyg}. This allows us to place a constraint on the initial PBH abundance, $\beta$.

The present-day DM abundance is $\Omega_{\text{DM,0}} = \rho_{\text{DM,0}} / \rho_{\text{crit},0}$. The DM energy density today, $\rho_{\text{DM,0}}$, is related to its energy density at the time of PBH evaporation, $\rho_{\text{DM}}^{\text{ev}}$, by the conservation of entropy ($s a^3 = \text{const}$) as
\begin{equation}
\rho_{\text{DM,0}} = \rho_{\text{DM}}^{\text{ev}} \left( \frac{g_{*s}(T_0) T_0^3}{g_{*s}(T_{\text{ev}}) T_{\text{ev}}^3} \right) ,
\end{equation}
where $T_0$ and $T_{\rm ev}$ are the radiation temperature at present and at the time of PBH evaporation respectively, and $g_{*s}(T)$ is the entropic degrees of freedom at temperature $T$. The DM energy density at evaporation is simply the number density of PBHs at that time, $n_{\text{BH}}^{\text{ev}}$, multiplied by the total number of DM particles produced per PBH, $N_{\text{DM}}$, and the DM mass, $m_{\text{DM}}$
\begin{equation}
\rho_{\text{DM}}^{\text{ev}} = n_{\text{BH}}^{\text{ev}} N_{\text{DM}} m_{\text{DM}} .
\end{equation}
Using the conservation of PBH number ($n_{\text{BH}}^{\text{ev}} (a_{\text{ev}})^3 = n_{\text{BH}}^{\text{in}} (a_{\text{in}})^3$) and entropy again, we can relate the present DM abundance to the \textit{initial} PBH number density $\rho_{\text{PBH}}^{\text{in}} = n_{\text{BH}}^{\text{in}} M_{\text{in}}$ as
\begin{equation}
\label{eq:omega_dm_intermediate}
\Omega_{\text{DM,0}} = \left( \frac{N_{\text{DM}} m_{\text{DM}}}{M_{\text{in}}} \right) \left( \frac{g_{*s}(T_0) T_0^3}{\rho_{\text{crit},0}} \right) \left( \frac{\rho_{\text{PBH}}^{\text{in}}}{g_{*s}(T_{\text{in}}) T_{\text{in}}^3} \right).
\end{equation}
To solve Eq.~(\ref{eq:omega_dm_intermediate}), we must express the initial PBH density $\rho_{\text{PBH}}^{\text{in}}$ and temperature $T_{\text{in}}$ in terms of the initial PBH mass $M_{\text{in}}$ and the initial abundance parameter $\beta = \rho_{\text{PBH}}^{\text{in}} / \rho_{\text{total}}^{\text{in}}$. Using the standard Friedmann equation and the definition of the initial PBH mass as a fraction $\gamma$ of the horizon mass, yields the final, practical formula for the present DM abundance to be~\cite{Haque:2024eyh, Cheek:2021odj}
\begin{equation}
\label{eq:omega_dm_final}
\Omega_{\text{DM,0}}h^2 \simeq 1.724 \left(\frac{\gamma}{0.2} \right)^{\frac{1}{2}} \left(\frac{g_{*}(T_{\text{in}})}{106.75} \right)^{-\frac{1}{4}} \left(\frac{1\text{g}}{M_{\text{in}}} \right)^{\frac{3}{2}} \left(\frac{m_{\text{DM}}}{1\text{GeV}} \right) \beta N_{\text{DM}} .
\end{equation}
This is the key equation for our constraint. To use it, we must first compute the total number of DM particles produced per PBH, $N_{\text{DM}}$ using Eq.~\eqref{total_N_i}. By setting $\Omega_{\text{DM,0}}h^2 = 0.12$, we can solve Eq.~(\ref{eq:omega_dm_final}) for $\beta$, giving the maximum allowed initial abundance of PBHs as a function of their initial mass $M_{\text{in}}$.

This procedure yields two distinct upper bounds on $\beta$, depending on whether the emitted particle mass $\mu_i$ is below or above the PBH's maximum post-accretion temperature $T_{\rm BH}^{\rm acc}$.

\noindent
\textbf{Case I: Continuous Emission ($\mu_i \lesssim T_{\rm BH}^{\rm acc}$)}

In the limit where the produced particle mass is below the maximum PBH temperature post-accretion, the integrated particle number yield $N_i$ remains proportional to $M_{\rm acc}^2$. Substituting the expression for $N_i$ into the constraint equation, the upper bound on the initial abundance $\beta$ is found to be
\begin{equation}
\beta \lesssim 9 \times 10^{-10} \left( 1-\frac{3 \lambda_{\rm Kerr}(a_*^{\rm acc}) \gamma}{8} \right)^2 \left( \frac{M_{\rm in}}{1 \rm g} \right)^{-1/2} \left( \frac{1 \rm GeV}{\mu_i} \right) \times
\begin{cases}
0.81 & s=0 \\
1.11 & s=1/2 \\
2.44 & s=1 \\
15.38 & s=2
\end{cases}.
\end{equation}

\noindent
\textbf{Case II: Delayed/Cut-off Emission ($\mu_i > T_{\rm BH}^{\rm acc}$)}

Conversely, when the particle mass exceeds the maximum temperature reached after accretion, the emission time window is cut short. This results in a constraint on $\beta$ that has a fundamentally different dependence on $M_{\rm in}$ and $\mu_i$. The corresponding maximum initial abundance is given by
\begin{equation}
\beta \lesssim 7 \times 10^{-36} \left( \frac{M_{\rm in}}{1 \rm g} \right)^{3/2} \left( \frac{1 \rm GeV}{\mu_i} \right)^{-1} \times
\begin{cases}
0.81 & s=0 \\
1.11 & s=1/2 \\
2.44 & s=1 \\
15.38 & s=2
\end{cases}.
\end{equation}
These two constraints govern the allowed parameter space in the $(M_{\rm in}, \beta)$ plane, defining the mass range where PBHs can account for the entirety of the observed dark matter density without overproducing relativistic particle species.

\begin{figure}[t]
\centering
\includegraphics[width=.499\textwidth]{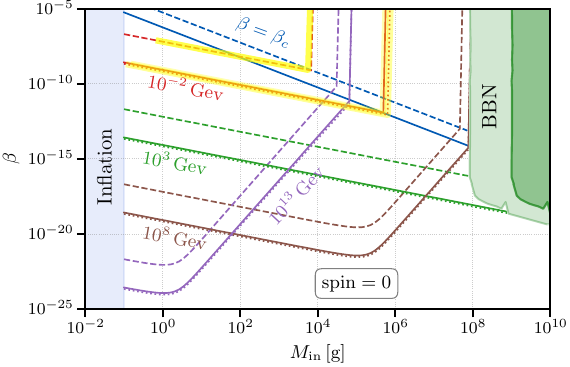}
\hspace{-0.3cm}
\includegraphics[width=.499\textwidth]{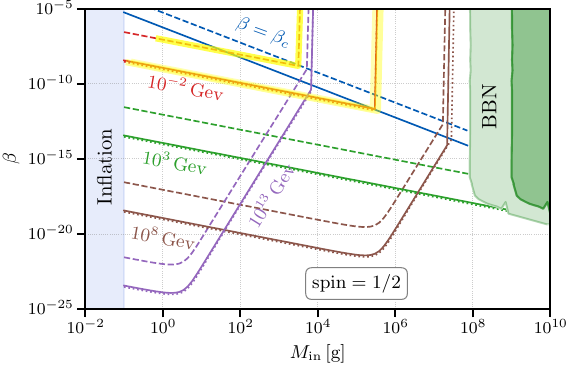}

\vspace{1em}

\includegraphics[width=.499\textwidth]{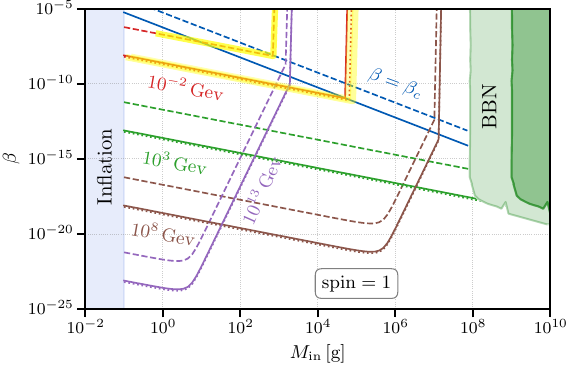}
\hspace{-0.3cm}
\includegraphics[width=.499\textwidth]{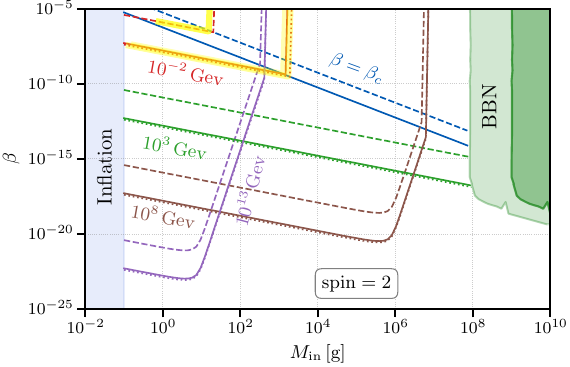}
\caption{Constraints on the initial PBH abundance fraction ($\beta$) versus initial PBH mass ($M_{\text{in}}$), shown for four different DM particle spins: $s=0$ (top-left), $s=1/2$ (top-right), $s=1$ (bottom-left), and $s=2$ (bottom-right). The colored curves trace the parameter space where PBH evaporation produces the observed DM relic density ($\Omega_{\text{DM,0}}h^2 \approx 0.12$) for the labeled DM masses ($m_{\rm DM}$). The line styles compare the standard evaporation-only scenario (dashed lines) with our full model that includes relativistic accretion for initial PBH spins of $a_* = 0.99$ (solid lines) and $a_* = 0.01$ (dotted lines). Shaded regions represent constraints from inflation (light blue), BBN (green), and improved BBN (light green). The yellow line indicates Ly-$\alpha$ limits.}
\label{fig:DM}
\end{figure}

\subsection*{Lyman-Alpha Constraint}

In addition to the constraint on the \textit{total abundance} of DM, a separate and powerful constraint comes from the \textit{kinematics} of the DM particles. Hawking radiation emits particles with energies characteristic of the BH's temperature, $E \sim T_{\text{BH}}$. For the PBH masses we are considering, this energy is typically much higher than the DM particle's rest mass, $m_{\rm DM}$. These DM particles are therefore produced as highly relativistic radiation. If these particles remain relativistic for too long, their large free-streaming length suppresses the formation of small-scale structures in the universe. This is in direct conflict with observations of the Lyman-alpha forest, which confirm the existence of structure on small scales~\cite{Murgia:2019duy, Saha:2024ies, Villasenor:2022aiy}.

This ``Warm Dark Matter'' (WDM) bound sets a lower limit on the DM particle's mass, $m_{\rm DM}$, as a function of the PBH mass that produced it. Following the analysis in \cite{Haque:2023awl, Barman:2024iht}, this constraint is given by
\begin{equation}
\label{eq:lyman_alpha}
\frac{m_{\rm DM}}{\text{GeV}} \geq 8.1 \times 10^7 \left(\frac{m_{\text{WDM}}}{\text{keV}} \right)^{\frac{4}{3}} \left(\frac{M_{\rm acc}}{M_p}\right)^{\frac{1}{2}} .
\end{equation}
Here, $m_{\text{WDM}}$ is the mass of a \textit{thermally} produced WDM particle that would have an equivalent free-streaming length. We adopt the conservative observational lower limit $m_{\text{WDM}} > 3.3 \text{ keV}$ \cite{Viel:2013fqw, Bode:2000gq}.

The PBH mass $M_{\rm acc}$ in the equation is the mass during the evaporation phase. In our model, this is not the initial mass $M_{\text{in}}$, but rather the significantly larger \textbf{accreted mass} $M_{\text{acc}}$ (from Eq.~\eqref{eq:mass_final_acc}). Since $M_{\text{acc}}$ is a strong function of the initial spin $a_*$, this Lyman-alpha constraint now becomes spin-dependent. A lower spin $a_*$ leads to a larger $M_{\text{acc}}$, which in turn imposes a \textit{stronger} (higher) lower bound on the DM particle's mass $m_{\rm DM}$.

To obtain the most accurate cosmological constraints, our calculation of the particle production rates ($\epsilon_i$, $\gamma_i$, and $\psi_i$) must use the complete, frequency-dependent greybody factors, which go beyond the simple geometric optics approximation. In this work, we perform these calculations using the publicly available Python package \textbf{\texttt{FRISBHEE}}\footnote{Available at \href{https://github.com/yfperezg/frisbhee/tree/58d4848aae5be4a347256a7e8d30e030493743d0}{https://github.com/yfperezg/frisbhee}.}~\cite{Cheek:2021odj, Cheek:2021cfe, Cheek:2022dbx, Cheek:2022mmy}. This allows us to accurately compute the total number of DM particles produced, $N_{\text{DM}}$, for the different DM masses and spins presented in our final results.

We present our final constraints on the PBH parameter space, combining our complete mass evolution model with the DM abundance calculation (Eq.~\ref{eq:omega_dm_final}) in Fig.~\ref{fig:DM}. The four panels display the required initial PBH abundance $\beta$ to explain 100\% of the DM as a function of the initial PBH mass $M_{\text{in}}$, for DM spins $s=0, 1/2, 1,$ and $2$. The key takeaway is the dramatic difference between the standard, evaporation-only model (dashed lines) and our new model (dotted and solid lines). Including relativistic accretion (dotted and solid lines) significantly strengthens the constraints, shifting the required $\beta$ to much lower values. This is because accretion extends the PBH's life, allowing it to produce a much larger total number of DM particles ($N_{\text{DM}}$) before evaporating. Therefore, a much smaller initial population ($\beta$) is needed to match the observed $\Omega_{\text{DM,0}}$. The strength of this new constraint is highly spin-dependent. The low-spin case ($a_* = 0.01$, dotted lines) shows the strongest effect, as these PBHs accrete most efficiently. The high-spin case ($a_* = 0.99$, solid lines) suppresses accretion, leading to a constraint that is closer to (but still much stronger than) the standard evaporation-only scenario. For all DM spins, our model shows that the inclusion of relativistic accretion rules out regions of the parameter space that were previously thought to be viable from BBN (light-green). We also plot the standard BBN constraint (green), the Lyman-$\alpha$ (WDM) constraint (yellow), which together carve out the allowed parameter space.

\section{Constraints on Evaporating PBHs as Dark Matter}\label{sec:f_PBH}

We now shift our focus from PBHs as a \textit{source} of DM to PBHs \textit{as} DM. In this scenario, we consider PBHs that survive until the present epoch and whose abundance is constrained by a variety of astrophysical observations.

\subsection*{The New Evaporation Threshold}

\begin{figure}[t]
\centering
\includegraphics[scale=1.05]{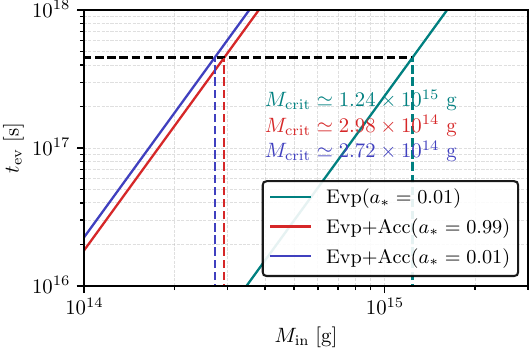} 
\caption{Evaporation time $t_{\mathrm{ev}}$ of PBHs as a function of their initial mass $M_{\mathrm{in}}$. The horizontal dashed line indicates the present age of the Universe, $t_0 \simeq 4.53\times10^{17}\,\mathrm{s}$. The intersection of each curve with this line determines the critical initial mass $M_{\mathrm{crit}}$ for PBHs surviving until today. Three cases are shown: the standard evaporation-only model (Evp, cyan), yielding $M_{\mathrm{crit}} \simeq 1.24\times10^{15}\,\mathrm{g}$; the evaporation plus accretion case for a high-spin PBH ($a_*=0.99$, red), giving $M_{\mathrm{crit}} \simeq 2.98\times10^{14}\,\mathrm{g}$; and for a low-spin PBH ($a_* = 0.01$, blue), giving $M_{\mathrm{crit}}\simeq 2.72\times10^{14}\,\mathrm{g}$.}
\label{fig:t_ev_vs_mi}
\end{figure}

In the standard, evaporation-only scenario, a PBH must have an initial mass $M_{\text{in}} \gtrsim  10^{15} \text{ g}$ to survive for the entire age of the universe~\cite{Carr:2009jm, Carr:2020xqk}. Any PBH with a mass below this limit is assumed to have evaporated, leaving no population today.

Our model, which includes relativistic accretion, fundamentally changes this premise. As shown in our mass evolution results (e.g., left panel of Fig.~\ref{total_mass_spin_evo}), a PBH with an initial mass can accrete to a significantly larger mass. This mass gain allows it to survive to the present day. The inclusion of accretion thus introduces a new, spin-dependent mass threshold. To re-evaluate this, we numerically solve our full evolution equations (Eqs.~\ref{eq:final_dmdt} and \ref{eq:final_dadt}) and determine the total evaporation time $t_{\text{ev}}$ for a given initial mass $M_{\text{in}}$. The results are shown in Fig.~\ref{fig:t_ev_vs_mi}. The horizontal dashed line represents the age of the universe, $t_0 \simeq 4.53 \times 10^{17} \text{s}$~\cite{Planck:2018vyg}. The intersection of any model with this line defines its \textit{critical mass}, $M_{\text{crit}}$, required for survival.
\begin{enumerate}
\item \textbf{Evaporation-Only:} The cyan curve shows the standard scenario, neglecting accretion. This model predicts a critical mass of $M_{\text{crit}} \approx 1.24 \times 10^{15} \, \text{g}$.
\item \textbf{Accretion $+$ Evaporation:} Our model, which includes relativistic accretion, yields a dramatically different result. New critial mass, therefore, can be obtained by assuming the fact that $\tau_{\rm BH} \simeq t_0$,
\begin{equation}
      M_{\rm crit} \simeq \left( 3 \epsilon M_p^4 t_0 \right)^{\frac{1}{3}} \left( 1 - \frac{3\lambda_{\text{Kerr}}(a_*^{\rm acc}) \gamma}{8} \right)  \simeq 1.2 \times 10^{15} \rm g  \left( 1 - \frac{3\lambda_{\text{Kerr}}(a_*^{\rm acc}) \gamma}{8} \right) .
\end{equation}
The blue ($a_*=0.01$) and red ($a_*=0.99$) curves show that for any given $M_{\text{in}}$, the PBH's lifetime is significantly extended. This is because the PBH first accretes to a much larger mass $M_{\text{acc}}$, and the evaporation timescale is extremely sensitive to this mass ($\tau_{\text{evap}} \propto M^3$). This effect lowers the critical initial mass required for survival by a factor of $\sim 4-5$.
\end{enumerate}
We also observe a clear spin dependence. The low-spin PBH (blue curve) accretes more efficiently, reaches a larger $M_{\text{acc}}$, and thus lives longer. This results in the lowest critical mass, $M_{\text{crit}} \simeq 2.72 \times 10^{14} \text{ g}$. The high-spin PBH (red curve) has its accretion suppressed, leading to a slightly shorter life and a critical mass of $M_{\text{crit}} \simeq 2.98 \times 10^{14} \text{ g}$. This finding is a key result of our paper. This naturally shifts the entire parameter space for the present-day PBH abundance, which is parameterized by $f_{\text{PBH}}(M_{\text{in}}) = \Omega_{\text{PBH}}(M_{\text{in}}) / \Omega_{\text{DM}}$, as a function of their initial mass $M_{\text{in}}$. Here, $\Omega_{\rm PBH}$ and $\Omega_{\rm DM}$ denote the present density parameters of PBHs and total DM, respectively.

\subsection*{Observational Constraints}

\begin{figure}[t]
\centering
\includegraphics[scale=1.30]{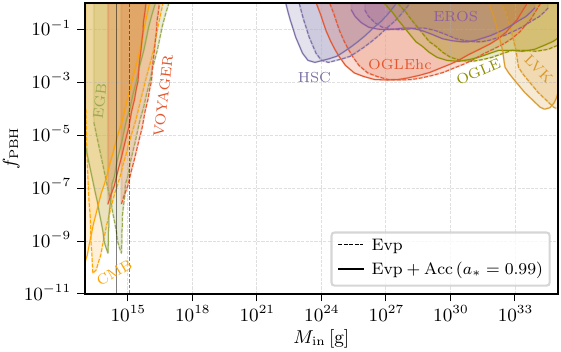} 
\caption{The final constrained parameter space for the present-day PBH dark matter abundance, $f_{\text{PBH}}$, as a function of the initial PBH mass, $M_{\text{in}}$. The shaded regions represent observational constraints (from Evaporation signatures, Microlensing surveys, and GW merger rates), which are applied to the PBH's initial mass, $M_{\rm in}$. Dashed lines represent the standard evaporation-only case. Our model (solid, $a_* = 0.99$) demonstrates that relativistic accretion shifts the entire parameter space to the left.
}
\label{fig:fPBH_vs_mi}
\end{figure}

The constraints on $f_{\text{PBH}}$ arise from a wide range of observational channels~\cite{Oncins:2022ydg,Carr:2016drx,Green:2020jor,Carr:2020xqk, Thoss:2024hsr}, which are sensitive to the PBH's \textit{present-day} mass. The key constraints include:
\begin{itemize}
    \item \textbf{Evaporation Signatures:} PBHs with masses above $10^{13}\,\mathrm{g}$ are constrained by CMB observations, since their evaporation during the recombination epoch would inject high-energy charged particles, thereby suppressing the CMB temperature anisotropies~\cite{Acharya:2020jbv}. For PBHs evaporating at later times in the low-redshift universe, even stronger limits arise from measurements of the extragalactic photon background (EGB)~\cite{Carr:2009jm} and from observations of the electron--positron spectra by Voyager~\cite{Boudaud:2018hqb}. These bounds probe PBH masses up to $ \sim 10^{17}\,\mathrm{g} $~\cite{Carr:2016hva, Korwar:2023kpy, Huang:2024xap}.
    
    
    \item \textbf{Gravitational Lensing:} Microlensing surveys (like OGLE, Kepler, EROS, and HSC) constrain the abundance of PBHs in the approximate range $10^{-10} M_{\odot}$--$10^{3} M_{\odot}$~\cite{Niikura:2019kqi, Niikura:2017zjd, Green:2016xgy}.
    \item \textbf{Gravitational Waves:} The merger rates of stellar-mass BHs detected by LIGO/ Virgo/ KAGRA place upper limits on the abundance of PBHs in that mass window~\cite{Andres-Carcasona:2024wqk, Nitz:2021vqh, Kavanagh:2018ggo, Vaskonen:2019jpv}.
\end{itemize}
Since our model shows a noticable shift in initial mass $M_{\text{in}}$ depending on its spin $a_*$, each of these constraints will be modified, leading to a noticeable change in the viable parameter space for PBH DM. We now present the final and most important result of our work in Fig.~\ref{fig:fPBH_vs_mi}. This plot shows the complete parameter space for the PBH as DM fraction $f_{\text{PBH}}$ as a function of the initial PBH mass $M_{\text{in}}$. The shaded regions are the observational constraints which are all fundamentally constraints on the present-day mass.

To prevent any ambiguity regarding the parameters shown in Fig.~\ref{fig:fPBH_vs_mi}, we explicitly clarify the definitions of the axes. The y-axis displays the present-day dark matter fraction, $f_{\rm PBH}$. Because the comoving number density of PBHs is conserved during the rapid early accretion phase, the total PBH energy density is naturally amplified by the mass ratio $M_{\rm acc}/M_{\rm in}$. Therefore, $f_{\rm PBH}$ represents the final, post-accretion abundance. This observable late-time abundance is plotted against the pre-accretion formation mass $M_{\rm in}$ on the x-axis. This mixed representation intentionally bridges the two epochs, providing a direct translation from late-time phenomenological constraints back to the initial parameter space required by early-universe formation models.

It is important to note a caveat regarding the mapping of these constraints. The shaded regions in Fig.~\ref{fig:fPBH_vs_mi} represent the most up-to-date present-day bounds~\cite{Carr:2026hot}. By plotting these against our initial mass $M_{\rm in}$, we are illustrating the leading-order effect of relativistic accretion: the translation of the mass bounds to significantly smaller initial formation masses. However, as detailed in recent literature~\cite{Carr:2026hot}, different constraints scale differently with mass and abundance. This is particularly true for constraints derived from PBH merger rates (GWs), which exhibit a highly non-linear dependence on both $M$ and $f_{\rm PBH}$. Therefore, while a horizontal shift effectively captures the primary consequence of mass growth, a fully precise recasting of specific bounds, such as the modification of the exact slope or amplitude of the GW constraint curve, requires recalculating the underlying physical rates using the accreted mass spectrum, which introduces secondary shape corrections to the shifted boundaries shown here. 

This result fundamentally alters the viable parameter space for PBH DM, demonstrating that relativistic accretion in the early universe is a critical and previously overlooked component.


\section{Stochastic Gravitational Wave Background}\label{sec:GW}

Evaporating PBHs emit gravitons via Hawking radiation, contributing to the SGWB. Unlike electromagnetic radiation, which is thermalized by the ambient plasma in the early universe, these gravitons decouple immediately and travel freely to the present epoch, carrying a ``snapshot'' of the PBH evaporation history.


The instantaneous emission rate of gravitons (spin $s=2$) from a single Kerr BH, per unit frequency interval, is given by~\cite{PhysRevD.14.3260}
\begin{equation}
\frac{d^2 N_{\text{grav}}}{dt d\omega} = \sum_{l,m} \frac{\Gamma^{lm}_{(s=2)}(\omega)}{e^{(\omega - m\Omega)/T_{\text{BH}}} - 1} \equiv Q_{\text{GW}}(t, \omega) .
\end{equation}
To compute this quantity accurately, we utilize the public code \texttt{BlackHawk}\footnote{Available at \href{https://blackhawk.hepforge.org}{https://blackhawk.hepforge.org}}~\cite{Arbey:2019mbc, Arbey:2021mbl}. This code numerically calculates the greybody factors and sums over the angular momentum modes $(l,m)$ to provide the precise graviton spectral rate $Q_{\text{GW}}(t, \omega)$ as a function of the instantaneous PBH mass $M(t)$ and spin $a_*(t)$.

\subsection*{Present-Day GW Abundance}

The differential energy density of GWs emitted by a PBH population with number density $ n_{\rm BH}(t) $ is
\begin{equation}
\frac{d^2\rho_{\rm GW}}{dt d\omega} =  \frac{\omega}{2\pi} n_{\rm BH}(t) Q_{\rm GW}(t, \omega),
\end{equation}
To obtain the present--day GW abundance, one must integrate the emission over the entire PBH lifetime and include the cosmological redshifting of
both frequency and energy density. The frequency redshifts as $\omega \propto a^{-1}$, and the GW energy density redshifts as $\rho_{\text{GW}} \propto a^{-4}$. The present-day GW energy density spectrum, normalized to the critical density $\rho_{\text{crit},0}$, is defined as~\cite{Anantua:2008am}
\begin{equation}
\Omega_{\text{GW}}(f_0) = \frac{1}{\rho_{\text{crit},0}} \frac{d\rho_{\text{GW}}^0}{d\ln f_0} .
\end{equation}
Integrating the redshifted flux from the time of formation $t_{\text{in}}$ to the time of complete evaporation $t_{\text{ev}}$, we obtain~\cite{Ireland:2023avg}
\begin{equation}
\Omega_{\text{GW}}(f_0) = \frac{n_{\text{BH}}^{\text{in}}}{ \rho_{\text{crit},0}} \frac{\omega_0^2}{2\pi} \frac{a^3_{\text{in}}}{a_0^2} \int_{t_{\text{in}}}^{t_{\text{ev}}} \frac{dt}{a(t)} \, Q_{\text{GW}}\left(t, \frac{\omega_0 a_0}{a(t)}\right) ,
\end{equation}
where $a_0 \equiv 1$ is the present--day scale factor and $\omega_0 = 2\pi f_0$ is the angular frequency observed today. It is convenient to express the initial PBH number density $n_{\text{BH}}^{\text{in}}$ in terms of the dimensionless abundance parameter $\beta$. Using the relation derived in previous sections, $n_{\text{BH}}^{\text{in}} = \beta \rho_{\text{total}}^{\text{in}} / M_{\text{in}}$, and substituting into the abundance equation, we arrive at our final expression for the SGWB spectrum
\begin{equation}\label{eq:omega_gw_final}
\Omega_{\rm GW}h^2 (f_0) = \frac{24 \pi \gamma^2 M_p^6 \omega_0^2}{\rho_{\rm crit,0}h^{-2} M_{\rm in}^3} \beta \, a_{\rm in}^3 \int_{a_{\rm in}}^{a_{\rm ev}} \frac{da}{H a^2} Q_{\rm GW}\left(M_{\rm BH}(a),a_*(a), \frac{\omega_0}{a}\right) .
\end{equation}

\subsection*{Numerical Implementation}

\begin{figure}
\centering
\includegraphics[scale=1.20]{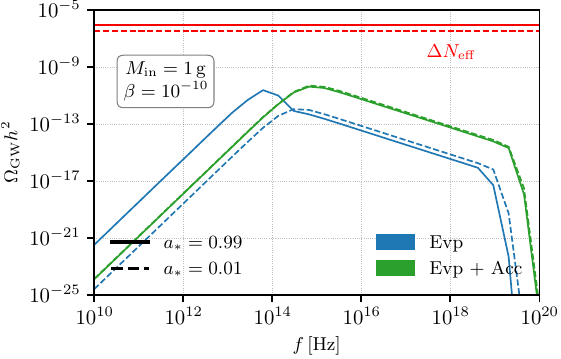} 
\caption{The present-day stochastic gravitational wave energy density spectrum, $\Omega_{\text{GW}}h^2$, as a function of frequency $f$, for a PBH population with initial mass $M_{\text{in}} = 1 \text{ g}$ and initial abundance $\beta = 10^{-10}$. We compare the standard evaporation-only scenario (Evp, blue curves) with our full model including relativistic accretion (Evp + Acc, green curves). The solid and dashed lines correspond to initial spins of $a_* = 0.99$ and $a_* = 0.01$, respectively.
The red horizontal lines indicate the bounds from the effective number of relativistic degrees of freedom, corresponding to $\Delta N_{\text{eff}} = 0.17$ (Planck + BAO)~\cite{Planck:2018vyg} and $\Delta N_{\text{eff}} = 0.06$ (CMB--S4)~\cite{Abazajian:2019eic}.}
\label{fig:GW}
\end{figure}

To evaluate Eq.~(\ref{eq:omega_gw_final}), we perform a consistent time-step integration. At each scale factor $a(t)$, we:
\begin{enumerate}
    \item Solve the coupled differential equations for accretion and evaporation (Eqs.~\ref{eq:final_dmdt} and \ref{eq:final_dadt}) to determine the instantaneous mass $M(a)$ and spin $a_*(a)$.
    \item Pass these values to the \texttt{BlackHawk} instantaneous spectrum script \verb|BlackHawk_inst| to compute $Q_{\text{GW}}$ at the blueshifted frequency $\omega = \omega_0 a_0 / a$.
    \item Integrate the resulting contribution over the expansion history of the universe.
\end{enumerate}
This procedure ensures that the effects of relativistic accretion—which modifies the mass evolution $M(t)$ and lifetime $t_{\text{ev}}$—are fully imprinted onto the predicted GW signal. We now present the resulting SGWB spectrum obtained from our numerical integration. Figure \ref{fig:GW} displays the dimensionless energy density $\Omega_{\text{GW}}h^2$ for a representative PBH population with initial mass $M_{\text{in}} = 1 \text{ g}$. The plot reveals profound differences between the standard evaporation scenario (Evp, blue) and our accretion-inclusive model (Evp + Acc, green).

The accretion process allows the PBH to grow to a mass $M_{\text{acc}} \gg M_{\text{in}}$. Since the Hawking temperature scales as $T_{\text{BH}} \propto 1/M$, the peak frequency of emission ($\omega_{\text{peak}} \sim T_{\text{BH}}$) shifts to lower values. Consequently, the green curves are shifted to the left compared to the blue curves. Furthermore, the extended lifetime and increased mass lead to a larger total energy release, enhancing the overall amplitude of $\Omega_{\text{GW}}$. A critical prediction of standard Hawking radiation is that high-spin BHs emit gravitons more efficiently at specific high-frequency modes due to superradiance and spin-dependent greybody factors. This creates a distinctive ``bump'' or secondary peak in the high-frequency tail of the spectrum, clearly visible in the standard high-spin case (solid blue line, $a_*=0.99$). However, when relativistic accretion is included (solid green line), this high-spin feature completely disappears. The spectrum for the initially high-spin PBH ($a_*=0.99$) becomes indistinguishable from the low-spin case ($a_*=0.01$). This is a direct consequence of the spin-down effect derived in Eq. (\ref{eq:final_dadt}). The accretion of zero-angular-momentum fluid during the early universe rapidly dilutes the PBH's spin. By the time the PBH enters its final, intense evaporation phase—where the bulk of the high-frequency gravitons are emitted—it has already spun down to $a_* \approx 0$. Therefore, regardless of their initial spin, accreting PBHs evaporate as Schwarzschild BHs, exhibiting a smooth, single-peak GW spectrum. This result implies that if a SGWB from light PBHs is detected in the future, the absence of high-spin spectral features could be a strong signature of an early accretion phase.

Note that our central findings and derived analytical approximations are anchored in the use of a monochromatic PBH mass function. To fully account for cosmological realism, however, one must consider an extended mass distribution $\psi(M_{\rm in})$. In this more generalized scenario, the total observable GW background is defined as the superposition of individual monochromatic spectra, weighted by the distribution function $\psi(M_{\rm in})$. Crucially, because the spectral peak frequency is directly mass-dependent \cite{Ireland:2023avg}, the effect of mass variance is to smear out the sharp features. Consequently, the distinct, singular peak corresponding to a single initial mass $M_{\rm in}$ is broadened across a range of frequencies dictated by the specific width and functional shape of the mass distribution itself.

\section{Conclusion}\label{sec:conclusion}

In this work, we have constructed a comprehensive and self-consistent framework for the evolution of spinning PBHs, from their formation in the radiation-dominated era to their eventual evaporation. We have demonstrated that a realistic model must simultaneously include the effects of general relativity, fluid dynamics, BH spin, and cosmology. Our primary contribution is the first-principles derivation of the relativistic accretion rate for a Kerr BH in an expanding universe, parameterized by a new, spin-dependent efficiency $\lambda_{\text{Kerr}}(a_*)$. We found that while relativistic effects enhance accretion compared to the standard Bondi-Hoyle model, this effect is significantly suppressed by the BH's spin, $a_*$. By combining this new accretion rate with the full, spin-dependent formalism for Hawking evaporation, we formulated a set of coupled evolution equations for the PBH mass $M(t)$ and spin $a_*(t)$. The numerical solution of these equations revealed two critical and previously overlooked physical effects:
\begin{enumerate}
    \item \textbf{Significant Mass Growth:} Relativistic accretion in the early universe is highly efficient (especially for low spin), causing a PBH's mass to grow $40-50\%$ from its initial formation mass ($M_{\text{acc}} \simeq 5 M_{\text{in}}$) before evaporation becomes the dominant process. This dramatically extends the PBH's lifetime, as $\tau_{\text{evap}} \propto\times (5 M_{\rm in})^3$.
    
    \item \textbf{Rapid Spin-Down:} The accretion of the zero-angular-momentum cosmic fluid acts as a powerful source of ``spin dilution'' ($\dot{a}_* \propto -a_* \dot{M}$). This effect is so strong that it rapidly spins down even a maximally rotating PBH ($a_* \sim 0.99$) to $a_* \approx 0$, long before the final evaporation phase.
\end{enumerate}

These two effects have profound and direct consequences for all major PBH constraints. Because accretion extends the PBH lifetime, the constraint from BBN becomes significantly stronger. We found that the upper limit on the initial PBH mass, $M_{\text{in}}$, is lowered by a factor of $\sim 4-5$. For the same reason, PBHs that produce DM via evaporation have a longer time to do so. This increases the total number of DM particles produced ($N_{\text{DM}}$) per PBH, which in turn strengthens the constraints on the initial PBH abundance, $\beta$. The accretion-induced mass growth fundamentally alters the mapping from the initial mass ($M_{\text{in}}$) to the present-day mass ($M_0$). We found that the critical initial mass required to survive until today is lowered from $M_{\text{in}} \sim 10^{15} \text{ g}$ to $M_{\text{in}} \sim 2.7 \times 10^{14} \text{ g}$. This shifts the entire $f_{\text{PBH}}$ parameter space to the left, mapping all observational constraints (Evaporation, Microlensing, GWs) onto a new, smaller range of initial masses. The rapid spin-down of the PBH erases the high-frequency ``bump'' in the SGWB spectrum, which was previously thought to be a smoking-gun signature of spinning PBHs. Our model predicts that all accreting PBHs, regardless of their initial spin, evaporate as Schwarzschild BHs, producing a smooth, single-peaked GW spectrum. In summary, we have shown that relativistic accretion in the early universe is not a small correction but a dominant effect in the evolution of PBHs. Any future study of PBH cosmology, constraints, or phenomenology must account for this critical phase of mass growth and spin-down to obtain realistic results.

\acknowledgments

Authors are very thankful to Prof. Santabrata Das, Subhankar Patra, and Gargi Sen for suggestions/clarification on various stages of the work on the important topic of accretion. The work of JK is supported by the Ministry of Human Resource Development, Government of India. 


\end{document}